Perpendicular-anisotropy artificial spin ice with spontaneous ordering: a platform for neuromorphic computing with flexible timescales

Aleksandr Kurenkov[1,2], Jonathan Maes[3], Aleksandra Pac[1,2], Gavin Martin Macauley[1,2], Bartel Van Waeyenberge[3], Aleš Hrabec[1,2] and Laura Jane Heyderman[1,2]

[1]Laboratory for Mesoscopic Systems, Department of Materials, ETH Zurich, 8093, Zurich, Switzerland
[2]Paul Scherrer Institute, Center for Neutron and Muon Sciences, Forschungsstrasse 111, 5232 Villigen, Switzerland
[3]DyNaMat, Department of Solid State Sciences, Ghent University, Belgium

ABSTRACT

Arrays of coupled nanomagnets have wide-ranging fundamental and practical applications in artificial spin ices, neuromorphic computing and spintronics. However, lacking in these fields are nanomagnets with perpendicular magnetic anisotropy with sufficient magnetostatic interaction. This would not only open up new possibilities for artificial spin ice geometries but also enable novel coupling methods for applications. Here we demonstrate a method to engineer the energy landscape of artificial spin lattices with perpendicular magnetic anisotropy. With this, we are able to realize for the first time magnetostatically-coupled 2D lattices of out-of-plane Ising spins that spontaneously order at room temperature on timescales that can be precisely engineered. We show how this property, together with straightforward electrical interfacing, make this system a promising platform for reservoir computing. Our results open the way to investigate the thermodynamics of out-of-plane magnetostatically coupled nanomagnet arrays with novel spin ice geometries, as well as to exploit such nanomagnet arrays in unconventional computing, taking advantage of the adjustable temporal dynamics and strong coupling between nanomagnets.

**<u>Introduction</u>**

Coupled nanomagnets organized on the sites of various lattices are widely-explored in the field of artificial spin ice[1–3] since they exhibit a large variety of fascinating phenomena including collective dynamics[4–6], frustration[7–9], dynamic chirality[10] and phase transitions[11–13]. Furthermore, these properties can be exploited for novel forms of conventional[14–19] and unconventional[20–24] computing. The single-domain nanomagnets typically have an in-plane magnetic anisotropy, which results in strong magnetostatic coupling between the magnets due to extended demagnetizing fields in the lattice plane. In contrast, dipolar-coupled nanomagnets with perpendicular magnetic anisotropy have not often been employed for artificial spin ice and have not shown spontaneous ordering – so they are not "thermally-active" – at experimentally measurable timescales[25–29].

Obtaining lattices comprised of magnetostatically-coupled nanomagnets with perpendicular anisotropy that show spontaneous ordering on an experimentally-feasible timescale would open the way to study the thermodynamics of a large variety of systems. Among them is the prototypical frustrated system, the triangular lattice with antiferromagnetic interactions[30], which has a rich magnetic phase diagram that depends on the details of the interactions among its component spins[31]. Crystal planes in many bulk magnetic systems can be approximated by triangular and other lattices of out-of-plane magnetic moments[32–34]. It would be beneficial to have an artificial spin ice equivalent to these bulk systems, since this would allow the magnetic configurations and magnetization dynamics to be visualised in real space with magnetic microscopy methods such as magnetic force microscopy (MFM)[6], Lorentz transmission electron microscopy[35] and synchrotron x-ray photoemission electron microscopy[5].

A particularly important model that can be constructed from out-of-plane spins is the canonical two-dimensional (2D) Ising model[36–38]. Being universally complete, so that any other statistical model[39,40] or Boolean circuit[41,42] can be derived from it, the 2D Ising model is of fundamental significance for statistical physics and has been used to model numerous physical, mathematical and biological processes[43–45]. The modelling is typically carried out by selecting a particular variant of the model, setting an initial state, and then observing how it evolves into a lower-energy state over time. The nature of this evolution provides a means to solve non-deterministic polynomial-time (NP-) hard problems with only polynomial overhead (P-hard) by mapping them onto a corresponding Ising system[46,47]. Since any NP-hard problem can be formulated as an Ising problem, the 2D Ising system is of great interest for a variety of combinatorial problems[44]. In addition, energy minimization and state dynamics in 2D Ising lattices bear similarities to equivalent processes in the human brain[48], providing a basis for computational models of the brain[48,49] and several types of neural networks such as Boltzmann machines[50] or Hopfield networks[51].

Here, we create for the first time a lattice of magnets with out-of-plane anisotropy that spontaneously orders at room temperature and in the absence of external magnetic fields. We focus on a 2D square lattice of Ising spins (Fig. 1a) because of its importance for fundamental science and for applications. By fabricating such lattices with nanomagnets with various diameters and separations, and made from precisely tuned multilayer films with different number and thicknesses of the layers, we have been able to determine both theoretically and experimentally how the energy landscape of the system defines its ordering dynamics. Utilizing this information, we have engineered the timescale of the system response, which provides a means to tailor the arrays for specific computing applications.

## Results

### Relating the energy landscape to the magnetic relaxation timescale with Monte Carlo simulations

We begin with a theoretical exploration of the degree and timescale of spontaneous ordering in a square 2D Ising system and how it is influenced by the energy landscape of the spins. For this, we performed kinetic Monte Carlo simulations using the "Hotspice" package[52]. The system simulated was a lattice of 11×11 magnetostatically (dipolarly, in this case) coupled Ising spins (Fig. 1a) in contact with a heat bath of temperature $T$. The lattice was first initialized in a state with all spins pointing up (upper panel of Fig. 1a) and allowed to relax for 1000 seconds. This time was chosen to match the approximate time between the initialization and measurement of the experimental system discussed later, with the average magnetization $m_{avg}$ and order parameter $q_{NN}$ tracked during this period. We chose a $q_{NN}$ given by $(1 - \langle S_i S_{i+1}\rangle)/2$, where $\langle S_i S_{i+1}\rangle$ is the nearest-neighbour correlation, and the mean switching time of a spin $j$ at a time $t$ is given by the Néel-Arrhenius law[53–55]:

$$\tau_j(t) = \tau_0 \exp\left(\frac{\Delta E_j(t)}{k_B T}\right). \qquad (1)$$

Here $t$ is elapsed time, $\tau_0 = 10^{-10}\ s$ is the attempt period and $\Delta E_j(t)$ is the energy barrier

$$\Delta E_j(t) = E_{EA}\big(1 + rand(j)\big) + E_{MC} \sum_{i,j} S_i(t) D_{ij} S_j(t), \qquad (2)$$

defined by a configuration of spins $S_i$ at every time step, magnitude of the point dipolar interaction $D_{ij}$ between spins $i$ and $j$, a random ± 5% Gaussian variation of anisotropy $rand(j)$ for different spins[24,56], as well as a time-independent effective anisotropy energy $E_{EA}$ and energy associated with the dipolar coupling between the nearest neighbour spins $E_{MC}$ (see Supplementary Information 1 for details).

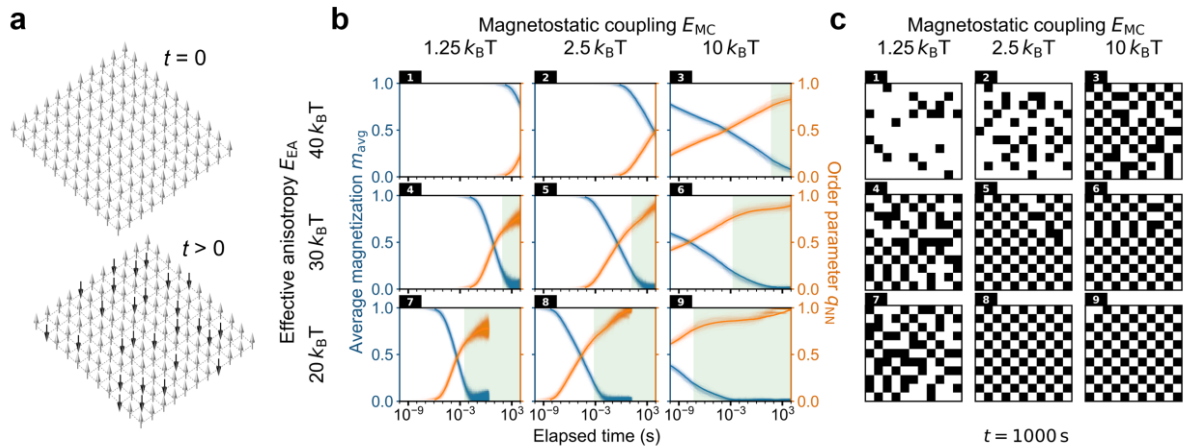

**Figure 1 | Dependence of the ordering timescale on the energy balance in a 2D Ising square lattice.** (**a**) Schematic of the system with 11×11 spins. It is initialized to a uniform magnetic state with all spins pointing up and then released at $t = 0$. The lattice then relaxes to lower energies for $t > 0$. This system size was chosen because it matched the size of our fabricated systems discussed later. (**b**) Evolution of the order parameter $q_{NN}$ and the average magnetization $m_{avg}$ for different effective anisotropy $E_{EA}$ and magnetostatic coupling $E_{MC}$. 20 simulations (fine lines) and their average (bold line) are shown for each case. The green shaded region highlights the second stage of ordering where the slope of $q_{NN}$ decreases and involves the reversal of spins within the domain boundary. The simulations are stopped after 1000 s and the resulting magnetic states are shown in panel (**c**). Here white (black) contrast corresponds to the up (down) spins.

The initial switching of individual nanomagnets is promoted by dipolar interactions with other nanomagnets in the lattice given by $E_{MC} \sum_{i,j} S_i(t) D_{ij} S_j(t)$ but, as the system assumes a more ordered state, this term decreases in value. This results in an exponential increase of the switching time due to the exponential term in Eq. 1. Simulations carried out for several sets of $E_{EA}$ and $E_{MC}$ confirm the universally logarithmic dependence of $m_{avg}$ and $q_{NN}$ on the elapsed time (Fig. 1b). An increase in $E_{MC}$ (see rows in Fig. 1b) extends the relaxation dynamics over a longer timescale, while changing $E_{EA}$ (see columns in Fig. 1b) does not change the slope of the curves and only shifts them along the time axis. Irrespective of the $E_{EA}$ and $E_{MC}$, the arrays achieve $m_{avg} \sim 0$ (blue traces in Fig. 1b) earlier than perfect checkerboard ordering with $q_{NN} = 1$ (orange traces in Fig. 1b). Furthermore, as the system nears $m_{avg} \sim 0$, the ordering rate decreases, which can be seen by the decrease in the slope of the orange lines as they enter the green-shaded regions in Figure 1b. Snapshots of the magnetic state at $t = 1000$ s (Fig. 1c) provide a clue to why this two-stage ordering process occurs (reflected by the two different slopes in $q_{NN}$). Panels 1 and 2 in Figure 1c are snapshots of the system during the rapid change of both $m_{avg}$ and $q_{NN}$ in the first phase of the ordering. Here, individual nanomagnets throughout the system switch (Panel 1) followed by domains of antiferromagnetic ordering starting to form (Panel 2). These doubly degenerate domains expand to completely fill the system (Panel 3 and 4) with domain boundaries forming between the domains with nanomagnets of opposite polarity (Panels 5 and 6).

From this point on, a further increase in $q_{NN}$ can only be achieved by switching the nanomagnets in the domain boundary. As these magnets are now stabilized through the dipolar interactions with their neighbours that have already switched, the energy barrier for them to switch is considerably higher than at the beginning of the relaxation process. This increases the switching time exponentially (see Eq. 1) and slows down further ordering significantly. The details of this process are elaborated in Supplementary Information 2. A consequence of the two-stage ordering is that, while a stronger coupling $E_{MC}$ promotes faster demagnetization $m_{avg} \rightarrow 0$ (e.g., see blue lines in Panels 7-9 of Fig. 1b, where the time to get to $m_{avg} \sim 0$ decreases from $\sim 10^{-2}$ to $\sim 10^{-3}$), it does not necessarily result in faster ordering $q_{NN} \rightarrow 1$ (orange lines in Panels 7-9 of Fig. 1b) as it increases the stabilization of the magnetic states in the domain boundaries. Therefore, $E_{MC}$ must not exceed a certain upper limit for the system to order on a given timescale and to not dwell in a metastable state with $m_{avg} \sim 0$ and $q_{NN} < 1$.

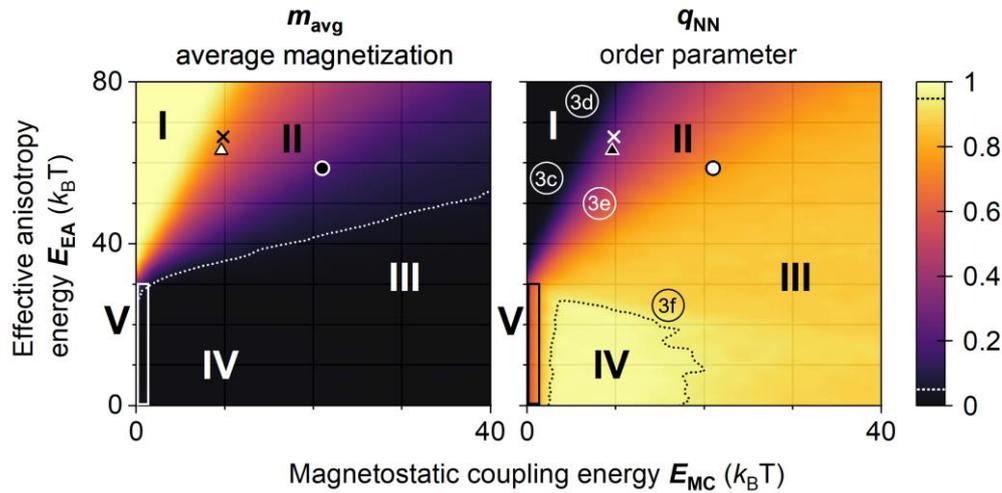

**Figure 2 | Phase diagrams of the average magnetization $m_{avg}$ and the order parameter $q_{NN}$ as a function of effective anisotropy $E_{EA}$ and magnetostatic coupling $E_{MC}$ at $t$ = 1000 s.** Five regions can be distinguished: I – frozen state, II – transient states, III – state with domains and domain boundaries, IV – checkerboard ordering, V – superparamagnetic state. The white (black) dotted line corresponds to $m_{avg}$ = 0.05 ($q_{NN}$ = 0.95). Labels 3c-3f correspond to the energy landscapes shown in Figure 3. The three symbols correspond to the experimental lattices with $D_{NM}$ = 170 nm, $t_{Co}$ = 1.45 nm and $S_{ASI}$ = 20 nm (circle), $S_{ASI}$ = 25 nm (triangle), $S_{ASI}$ = 30 nm (cross). The experimental magnetic configurations of these systems are shown in Figure 4.

The dependence of $m_{avg}$ and $q_{NN}$ on $E_{EA}$ and $E_{MC}$ at $t$ = 1000 s is summarized in the phase diagrams of Figure 2. In Region I, neither $m_{avg}$ nor $q_{NN}$ experience a significant change compared to the initial state due to the strong anisotropy and weak magnetostatic coupling (corresponding to Panel 1 in Fig. 1c). Increase of the $E_{MC}/E_{EA}$ ratio shifts the system though the transient Region II, characterised by decreasing average magnetization and increasing ordering (corresponding to Panel 2 in Fig. 1c). Further increase of $E_{MC}/E_{EA}$ shifts the system to Region III with $m_{avg} \sim 0$ and $q_{NN}$ still not 1, characteristic for the second ordering stage (corresponding to Panel 6 of Fig. 1c). Any further increase of $E_{MC}/E_{EA}$ does not affect the average magnetization or ordering at $t$ = 1000 s. Achieving the high ordering of Region IV, given by the dotted black line that envelops a region with $q_{NN} \geq 0.95$ in Figure 2 (exemplified by Panels 8 and 9 in Fig. 1c), requires lowering both $E_{MC}$ and $E_{EA}$ below a certain threshold. The narrow Region V (enclosed by vertical box in Fig. 2) is the regime in which the spins are superparamagnetic at the chosen timescale. The border between Regions V and I at $E_{MC}$ = 0 is therefore a transition between the superparamagnetic and frozen regimes. This border is located at $E_{EA} \sim 30$ $k_B$T because nanomagnets with such an effective anisotropy have an average switching time of $\sim 1000$ s (the time point of these diagrams). Region IV with $q_{NN} \sim 1$ is, therefore, constrained from the left by the superparamagnetic regime, from the top by the frozen regime, and from the right by states in which switching spins in domain boundaries is too energetically expensive.

The phase diagrams in Figure 2 are snapshots at $t$ = 1000 s and, with increasing time, the boundaries between the different regions will shift. Because of this, if the observation is long enough and $E_{MC}$ is non-negligible, a system located in the frozen region at $t$ = 1000 s may eventually find itself in the more ordered Regions II, III or IV (see phase diagrams for $t \sim 27$ months in Supplementary Information 3). The state of the array is therefore defined by $E_{EA}$, $E_{MC}$ and $t$. Practical applications exploiting systems with specific dynamics defined by $E_{EA}$ and $E_{MC}$ require careful adjustment of the lattice and nanomagnet dimensions, as well as the stack materials and layer thicknesses. We show how this can be achieved in the next section.

## Designing the energy landscape of coupled Co/Pt nanomagnets

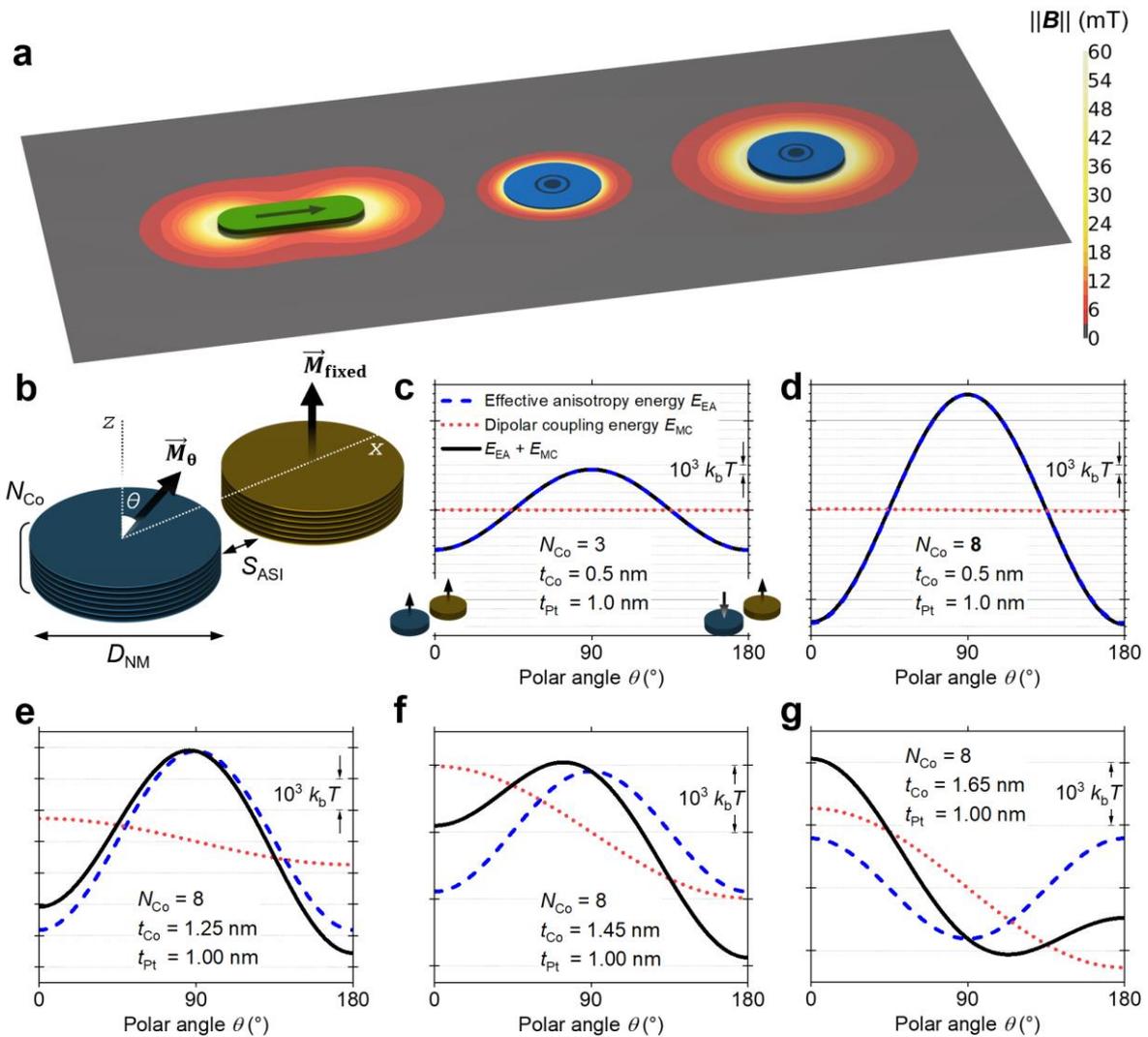

**Figure 3 | Dependence of the energy landscape of multilayered nanomagnets with perpendicular magnetic anisotropy on the material stack and lateral dimensions. (a)** Magnitude of the magnetic flux density for a permalloy nanomagnet (dimensions: 100×300×20 nm³) with in-plane anisotropy (left) and 200 nm-diameter nanomagnets with perpendicular

magnetic anisotropy consisting of 3 Co layers of 0.5 nm thickness (centre) and 8 Co layers of 1.45 nm thickness (right). The Co layers are separated by 0.8 nm of vacuum. (**b**) Schematic showing the geometrical parameters used in energy landscape calculations. (**c**)-(**g**) Energy landscapes calculated for $D_{NM}$ = 200 nm, $S_{ASI}$ = 20 nm and different Co layer thicknesses. The vertical scale is energy, where $k_B$ is Boltzmann constant and T = 300 K.

Artificial spin ices with perpendicular magnetic anisotropy that are thermally-active at room temperature on experimentally measurable timescales have not been observed before. The first reason for this is that achieving large $E_{MC}$ for this geometry is harder due to the confinement of demagnetizing fields in the vicinity of the nanomagnet (central magnet in Fig. 3a) compared to the in-plane case (leftmost magnet in Fig. 3a). The second reason is that balancing the energy contributions to $E_{EA}$, required to lower the energy barrier to switching, is equally challenging in Co/x (x = Pt, Pd, Ni) multilayers. To address these challenges, we focused on precisely controlling $E_{EA}$ in perpendicular nanomagnets and maximizing $E_{MC}$ between them.

We fabricated the nanomagnets from Co/Pt ferromagnetic multilayers, which are widely used because of their strong perpendicular magnetic anisotropy[57]. The primarily interfacial origin of the anisotropy allowed us to vary $E_{EA}$ and $E_{MC}$ almost independently by changing the number of Co/Pt interfaces and the thickness of the Co layers, respectively. We have calculated the energy landscape of a pair of nanomagnets to guide the selection of nanomagnet diameter $D_{NM}$, separation $S_{ASI}$, number of Co layer repetitions $N_{Co}$, and thicknesses of Co and Pt layers, $t_{Co}$ and $t_{Pt}$, respectively (Fig. 3b). We allowed the magnetization to rotate coherently in one nanomagnet ($\vec{M}_\theta$ in Fig. 3b) while keeping the other fixed ($\vec{M}_{fixed}$ in Fig. 3b). The energy was then calculated for polar angles $\theta$ from 0° to 180° and included four energy terms associated with the uniaxial interfacial anisotropy, demagnetization, magnetostatic interaction between the layers in the nanomagnet and the dipolar coupling between the nanomagnets (see Supplementary Information 4 for details). The resulting energy landscape provides an estimate of the energy barrier that the system needs to overcome to go from the higher-energy magnetic state with parallel moments to the lower-energy antiparallel state.

In Figure 3c, we show the energy landscape for a pair of coupled nanomagnets made of (Co [0.5] / Pt [1.0])$_2$ / Co [0.5] layers with $N_{Co}$ = 3, $D_{NM}$ = 200 nm and $S_{ASI}$ = 20 nm, where the numbers in square brackets are thicknesses in nm. Such stacks have a high anisotropy and are widely used for spintronics applications[58,59] due to the large thermal stability, which provides more than 10 years of retention time (a measure of how long a device can store information reliably without the nanomagnet switching) even in sub-20 nm diameter nanomagnets[58]. The small number of ferromagnetic Co layers, as well as their low thickness, results in a localized demagnetizing field (central magnet in Fig. 3a), minimizing crosstalk between the nanomagnets. Both properties–high anisotropy and low crosstalk–while useful for information storage applications, contradict the high-$E_{MC}$, low-$E_{EA}$ requirements of a lattice to give fast spontaneous ordering. Arrays of nanomagnets made of such a stack are located in the "frozen" Region I of Figure 2 ("3c" label).

To devise a high-$E_{MC}$, low-$E_{EA}$ stack, we varied $N_{Co}$ and $t_{Co}$ while fixing the nanomagnet diameter $D_{NM}$ and separation $S_{ASI}$. $S_{ASI}$ should be minimized to maximize the dipolar coupling, and was set to 20 nm, which was the smallest nanomagnet separation we were able to obtain with confidence when fabricating nanomagnet arrays with electron beam lithography. $D_{NM}$ was set to 200 nm, which was the largest possible nanomagnet diameter that did not result in the formation of multidomain states. This upper threshold value for $D_{NM}$ was determined experimentally by observing the magnetic states in multiple lattices of nanomagnets fabricated with different diameters.

Increasing the number of Co layer repetitions $N_{Co}$ from 3 to 8 produces a stack similar to those used in previous works on artificial spin ices with out-of-plane nanomagnets[25–29], with the energy landscape shown in Figure 3d. $E_{MC}$ is approximately quadratically proportional to the total Co thickness and increases by a factor of ~7. However, the increased number of Co/Pt interfaces results in a higher anisotropy, as seen in Figures 3c and 3d, and the system remains in Region I of Figure 2 ("3d" label). Achieving an $E_{MC}$-dominated lattice requires the lowering of $E_{EA}$ without decreasing $E_{MC}$. For this, one needs to consider the two additional components of $E_{EA}$ beyond the interfacial anisotropy: the demagnetization energy and dipolar coupling between layers of the stack. The dipolar coupling between the layers is rather weak and increases the perpendicular anisotropy, and thus cannot be used to lower $E_{EA}$. In contrast, the demagnetization energy is significant and decreases $E_{EA}$. The demagnetization energy increases with Co thickness, just like $E_{MC}$, which has the same physical origin. Therefore, increasing $t_{Co}$ to 1.25 nm results in a significant reduction in $E_{EA}$ and an increase in $E_{MC}$ (Fig. 3e), shifting the system towards Region II of Figure 2 ("3e" label). At $t_{Co}$ = 1.45 nm (Fig. 3f), $E_{MC} \sim E_{EA}$, which results in a highly asymmetric landscape with the lowest $E_{EA}$ and highest $E_{MC}$ among the considered stacks. Such an energy landscape should facilitate spontaneous ordering and, depending on the absolute values of $E_{EA}$ and $E_{MC}$, an array of such nanomagnets belongs to Region III or IV of Figure 2 ("3f" label). The demagnetizing field of these nanomagnets (rightmost nanomagnet in Fig. 3a) has a similar extent to that of the Permalloy nanomagnets typically implemented for in-plane artificial spin ices (leftmost nanomagnet in Fig. 3a). Further increase of $t_{Co}$ to 1.65 nm results in the shift of the energy minimum to $\theta \sim 110^{\circ}$, indicating a loss of perpendicular anisotropy (Fig. 3g).

From calculations, we have shown the effect on the energy landscape of changing the experimental system in terms of layer thicknesses, as well as the nanomagnet diameters and separations. By comparing this to the results from our Monte Carlo simulations, we are then able to predict how changing the system parameters will influence the relaxation timescale. This has enabled us to experimentally realize square lattices of interacting out-of-plane nanomagnets with the ability to control the ordering timescales from sub-seconds to years, as we show in the following section.

**Control of relaxation timescales in experimental dipolar-coupled 2D Ising systems**

We now turn to experimental systems where limitations in fabrication impose additional constraints on the system design. For example, achieving small $S_{ASI}$—a key parameter influencing $E_{MC}$—becomes more challenging as the thickness of the stack increases, so we limit $N_{Co}$ to 7. Another effect is that the Pt spacer layers may become discontinuous on decreasing their thickness below ~0.4 nm and, to be sure that we have a continuous layer we choose $t_{Pt}$ = 0.8 nm for the experimental systems, unless otherwise mentioned. These parameters permitted an $S_{ASI}$ down to 20 nm, and we varied $S_{ASI}$, $D_{NM}$ and $t_{Co}$ in the experiment.

The multilayers of Ta[3] / Pt[4] / (Co[$t_{Co}$] / Pt[0.8])$_6$ / Co[$t_{Co}$] / Ru[2] were deposited by magnetron sputtering on Si substrates with a natural oxide layer. They were patterned into square lattices of 11×11 nanomagnets by electron beam lithography and Ar ion milling. For transport measurements, the Ta/Pt bottom layer was patterned into Hall bars. Magnetic force microscopy (MFM) was used to measure the magnetic state after applying an out-of-plane magnetic field of 0.9 T to initialize all nanomagnets to an "up state". For the MFM measurements, we covered the samples with a 60 nm-thick Poly(methyl methacrylate) resist to minimize the influence of the stray magnetic field of the MFM probe on the sample.

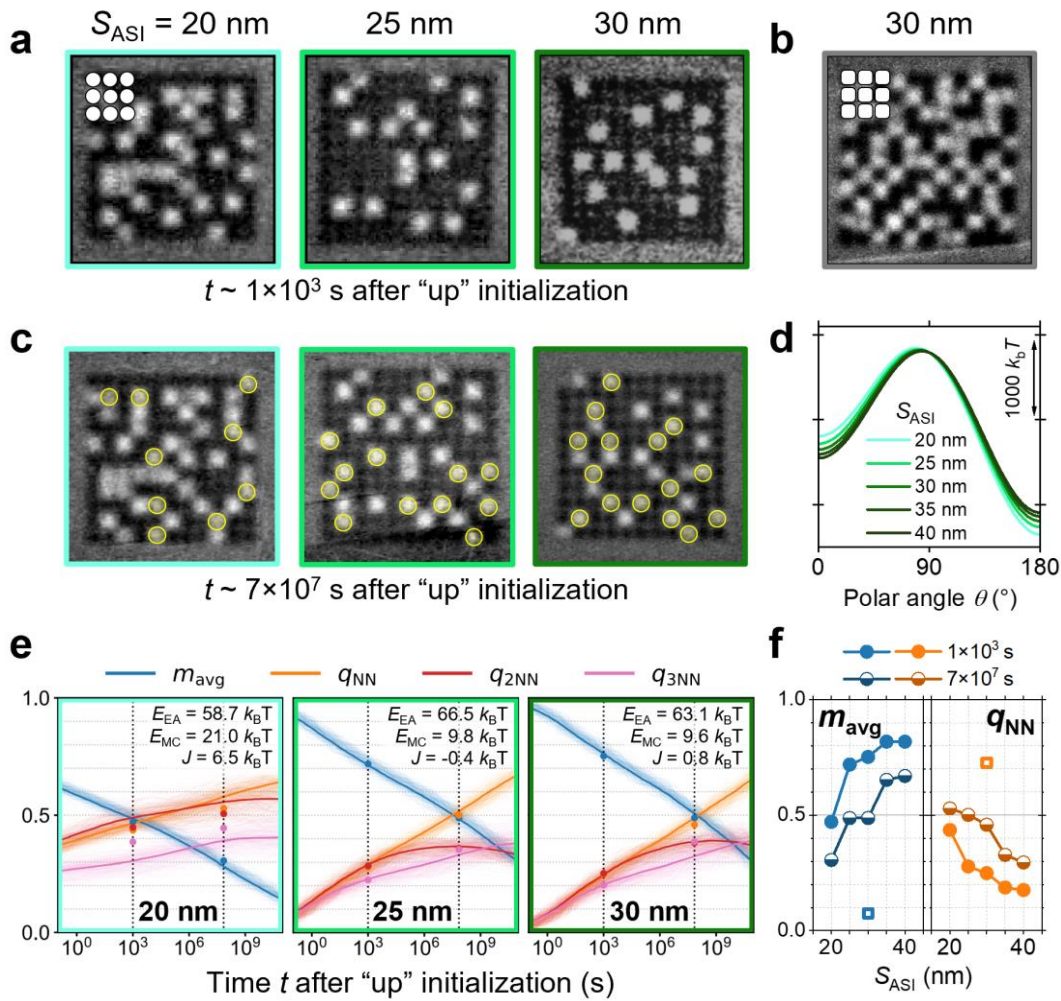

**Figure 4 | Observing the dependence of the ordering timescale on nanomagnet separation and nanomagnet shape with MFM.** All systems were initialized to a uniform magnetic state with an out-of-plane magnetic field before the measurements. (**a**) MFM images of lattices with $t_{Co}$ = 1.45 nm, $D_{NM}$ = 170 nm and increasing nanomagnet separation $S_{ASI}$ observed at $t$ ~ 1000 s after initialization. Black nanomagnets are in the initial magnetic state while white nanomagnets have switched. Image frame colours match those of the corresponding curves in (d) and frames in (e). (**b**) Array of 170 nm-wide square nanomagnets separated by 30 nm observed at $t$ ~ 1000 s. $t_{Co}$ = 1.45 nm as in (a). The nanomagnet shape is shown in the top left inset. All other results are for circular nanomagnets as shown in the top left inset of (a). (**c**) The same arrays as in (a) observed after ~27 months at room temperature. The magnets that have switched during this time are highlighted with yellow circles. (**d**) Energy landscapes calculated for the lattices shown in (a). (**e**) Evolution of average magnetization $m_{avg}$ and order parameters $q_{NN}$, $q_{2NN}$, $q_{3NN}$. The fine lines are calculated using Monte Carlo simulations, the bold lines are their averages, and the points are experimental data. (**f**) Average magnetization $m_{avg}$ and order parameter $q_{NN}$ as a function of $S_{ASI}$ determined from the MFM images. Data for panels (a), (b) and (c) are represented as filled circles, open squares and semi-filled circles, respectively. MFM images for $S_{ASI}$ = 35 and 40 nm can be found in Supplementary Information 5.

We first looked at arrays with $t_{Co}$ = 1.45 nm, $D_{NM}$ = 170 nm and different $S_{ASI}$. The state of the arrays at time $t \sim 1000$ s after initialization is shown in Figure 4a. Black contrast indicates that the nanomagnets are in the state initialized by magnetic field, while white contrast indicates nanomagnets that have spontaneously switched. The number of switched nanomagnets gradually decreases with increase in $S_{ASI}$. Since $S_{ASI}$ has no effect on $E_{EA}$ and only changes $E_{MC}$, this result is easy to interpret. The decrease of $E_{MC}$ at a constant $E_{EA}$ lowers the asymmetry of the energy landscape by making the initial energy well deeper and therefore harder to leave ($\theta = 0^0$ in Fig. 4d) while making the lower-energy well shallower ($\theta = 180^0$ in Fig. 4d).

From the MFM images, we calculated the average magnetization $m_{avg}$ and order parameter $q_{NN}$ as defined above. With decreasing $S_{ASI}$, the monotonic increase in $q_{NN}$ and decrease in $m_{avg}$ (filled circles in Fig. 4f; MFM images for 35 and 40 nm are shown in Supplementary Information 5) highlight the important role that the magnetostatic coupling plays in the spontaneous ordering. We do not reach $q_{NN}$ = 1, $m_{avg}$ = 0, characteristic of highly ordered states since a further decrease of $S_{ASI}$ is challenging in terms of the fabrication. Instead, we changed the shape of the nanomagnets from circular to square with rounded corners (compare insets in Fig. 4a and b) without modifying the stack. The resulting closer proximity of the magnets produces higher $E_{MC}$ and a significant increase in ordering (MFM in Fig. 4b; $m_{avg}$ and $q_{NN}$ shown as a square data point in Fig. 4f).

The $m_{avg}$ and $q_{NN}$ of a 2D Ising system with any $E_{EA}$ and $E_{MC}$ will evolve with time (see, for example, Fig. 1b). Therefore, one can achieve a higher degree of ordering simply by waiting long enough. To confirm this, we kept the arrays at room temperature in ambient atmosphere and no magnetic field for ~27 months, which is equivalent to ~$7 \times 10^7$ seconds. We then remeasured the samples without applying any magnetic field and observed the states shown in Figure 4c. The nanomagnets that have additionally switched are highlighted by yellow circles. There was no switching of nanomagnets from white to black (i.e. back to the state initialized 27 months prior). The new $m_{avg}$ and $q_{NN}$ are shown in Figure 4f with half-filled circles. The decrease in $m_{avg}$ and increase in $q_{NN}$ indicate the additional ordering of the lattice, and one can make use of the evolution of these parameters to pinpoint location of the lattices on the phase diagrams of Figure 2. For this, we calculated $m_{avg}$, $q_{NN}$ as well as the 2nd and 3rd nearest neighbour order parameters $q_{2NN}$ and $q_{3NN}$ from the MFM images for $t$ = 1000 s and $t$ = $7 \times 10^7$ s. We then fitted these values with $m_{avg}(t)$, $q_{NN}(t)$, $q_{2NN}(t)$ and $q_{3NN}(t)$ calculated using Monte Carlo simulations with the "Hotspice" package[52], and using $E_{EA}$, $E_{MC}$ and ferromagnetic exchange coupling between magnets $J$ as variables (see Supplementary Information 5 for details). The fitted results are shown in Figure 4e. The non-zero $J$ in the case of $S_{ASI}$ = 20 nm may indicate incomplete separation of the nanomagnets in the bottom Co layer. The near-zero $J$ for $S_{ASI}$ = 25 and 30 nm suggests that the magnets are fully separated. The location of the fitted results on the phase diagrams of Figure 2 (circles, triangles and crosses for $S_{ASI}$ = 20, 25 and 30 nm, respectively) indicates that the achieved levels of $E_{MC}$ are sufficient for a complete ordering (Region V) in $t \sim 1000$ s but $E_{EA}$ is too high. Note that these phase diagrams were calculated assuming that there is no exchange coupling between the nanomagnets, so $J$ = 0. Therefore, since the $S_{ASI}$ = 20 nm "circle" experimental point is for the lattice with $J \sim 6.5$ mT, its location is approximate.

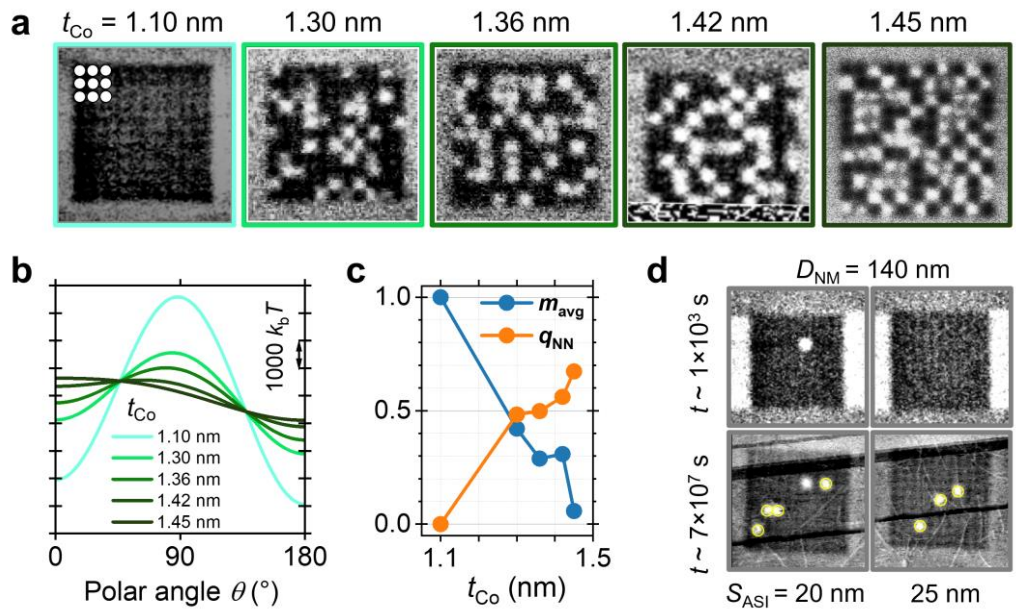

**Figure 5 | Dependence of ordering on the Co layer thickness and the nanomagnet size observed with MFM.** All systems were initialized to a uniform magnetic state (black contrast) with an out-of-plane magnetic field before the measurements. (**a**) MFM images of lattices with $D_{NM}$ = 200 nm, $S_{ASI}$ = 20 nm and increasing Co thickness $t_{Co}$ observed at $t \sim 1000$ s. Image frame colours correspond to those of the curves in (b). White contrast indicates that the nanomagnets have switched. (**b**) Energy landscapes calculated for the lattices shown in (a). (**c**) Average magnetization $m_{avg}$ and order parameter $q_{NN}$ as a function of $t_{Co}$ determined from MFM images in (a). (**d**) MFM images of lattices with $D_{NM}$ = 140 nm and $t_{Co}$ = 1.45 nm observed at $t \sim 1000$ s and $t \sim 7 \times 10^7$. The nanomagnets that have switched between the two times are indicated with a yellow frame.

Having looked at the effect of $S_{ASI}$ and time $t$ on ordering in the system, we then looked at the influence of $t_{Co}$ and $D_{NM}$. MFM images taken at $t \sim 1000$ s of the arrays with $D_{NM}$ = 200 nm, $S_{ASI}$ = 20 nm and varying $t_{Co}$ from 1.1 nm to 1.45 nm are shown in Figure 5a. As we have seen, $t_{Co}$ has a profound effect on $E_{EA}$ and $E_{MC}$ (Fig. 3e-g), and consequently on the ordering timescale. Indeed, as $t_{Co}$ increases, the corresponding energy landscapes become more asymmetric (Fig. 5b) and, from the MFM images, we find that $m_{avg}$ decreases while $q_{NN}$ increases (Fig. 5c). Increasing $D_{NM}$ to 200 nm provides a lower $m_{avg}$ and higher $q_{NN}$ (0.06 and 0.67, respectively; see data for $t_{Co}$ = 1.45 nm in Fig. 5c) than for the same lattice with $D_{NM}$ = 170 nm (0.47 and 0.44, respectively; see data for $S_{ASI}$ = 20 nm in Fig. 4f). However, further increase of $D_{NM}$ to 230 nm in the same stack results in formation of multidomain states within the nanomagnets, providing an upper limit to $D_{NM}$. Conversely, decrease of $D_{NM}$ to 140 nm slows down the ordering process so that there is hardly any switching at $t \sim 1000$ s (top images in Fig. 5d) and only a few reversed magnets at $t \sim 7 \times 10^7$ s (bottom images in Fig. 5d). We have therefore experimentally confirmed that $D_{NM}$ and $t_{Co}$ have a profound effect on the ordering dynamics of the lattice with $t_{Co}$ requiring a precise adjustment to achieve a low enough $E_{EA}$ (as shown in Fig. 3d-g). The out-of-plane anisotropy of the nanomagnets means that the state of the nanomagnets can be directly accessed with electrical readout using the anomalous Hall effect. To demonstrate this, we performed electrical measurements on lattices with $D_{NM}$ = 140 and 170 nm and $S_{ASI}$ = 30 and 40 nm on timescales of tens of seconds, which showed similar degrees of thermal activity to the lattices measured with MFM (Supplementary Information 6).

We have now demonstrated experimentally that the relaxation dynamics of out-of-plane spin lattices can be engineered with careful tuning of the anisotropy of and magnetostatic coupling between the nanomagnets by altering their dimensions, separation and shape, as well as the stack layer thicknesses. We have provided numerical evidence that the timescale associated with the collective dynamics can be tuned all the way from sub-seconds to years. This opens up new avenues for optimising the performance of neuromorphic computing systems incorporating these lattices as we show in the next section.

**Tuneable-frequency reservoir computing with 2D Ising systems**

The 2D Ising system is an appealing platform for a variety of computational approaches. Short-term memory and reset of the system after processing an input play a central role in brain-inspired computing[60,61], and both of these properties can be achieved by exploiting the dynamics of $m_{avg}$ and $q_{NN}$ that we presented above. This ability to tune the dynamics is particularly important in reservoir and probabilistic computing where, in contrast to conventional von Neumann computing, higher frequencies are not inherently beneficial. Instead, the priority lies in matching the dynamics of the physical system to the timescale of incoming data patterns since this is the only way that the physics of the system can be effectively harnessed for data processing. To demonstrate the effect of frequency matching on the computational performance, we have simulated a signal transformation task using our 2D Ising system for reservoir computing.

Reservoir computing is a framework for neural networks that makes use of a system with non-linear behaviour called a "reservoir" to map input data into a higher-dimensional space, in which the inputs can be separated by a linear transformation. The reservoir does not need to be trained (so it is not modified itself) but does need to be a system with short-term memory and high dimensionality[60,62], properties that are met by many physical systems[20,21,24,62–67]. Here we employ Monte Carlo simulations using the "Hotspice" package[52] to simulate the performance of a reservoir comprising the 2D Ising system in transforming sine waves of different frequencies into a sawtooth signal, a task that is often used to test the nonlinearity of a reservoir[67,68].

For the reservoir, we employed the 11×11 lattice with $E_{EA}$ = 20 $k_B$T ± 5% and $E_{MC}$ = 2.5$k_B$T (Panel 8 of Fig. 1b). We have experimentally demonstrated electrical readout of $m_{avg}$ in the 11×11 lattices using the anomalous Hall effect (see Supplementary Information 6), and therefore we implemented this magnetic state readout method in the simulations. The input for the simulations was applied in the form of an out-of-plane magnetic field $B(t)$ acting on the entire system (Fig. 6a) and was scaled such that its magnitude extends from $B_0$ (minimum) to $B_1$ (maximum) as shown in the top panel of Figure 6c. Accordingly, the energy barrier term of Equation 2 was extended to include the Zeeman term associated with the applied magnetic field as follows:

$$\Delta E_j(t) = E_{EA}\big(1 + rand(j)\big) + E_{MC}\sum_{i,j} S_i(t)D_{ij}S_j(t) + E_{Zeeman}(B(t), j). \qquad (3)$$

The average magnetization of each lattice column was used for the readout $y_l(t)$ (Fig. 6a). A discussion about the experimental feasibility of such a grid of local readouts is given in Supplementary Information 7. The readout was used to perform a linear regression $o(t) = \sum_{i=1}^{n} w_i\, y_i(t)$ and transformation[69,70], where $i$ is the array column number, $w_i$ is the associated weight, $o(t)$ is the predicted signal and $n$ = 11 is the array dimension. The inverse mean squared error (1/MSE) between $o(t)$ and the desired result (the sawtooth) was used as a performance metric.

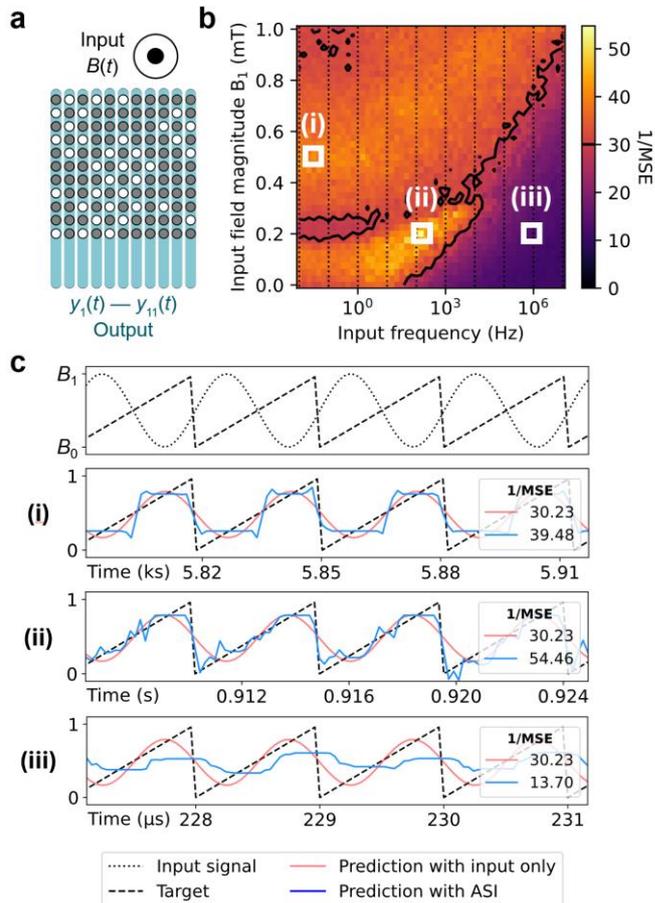

**Figure 6 | Monte Carlo simulations of the performance of an 11×11 lattice reservoir for a sine to sawtooth signal transformation.** The parameters used for the simulations are $E_{MC}$ = 2.5 $k_B$T, $E_{EA}$ = 20 $k_B$T ± 5% and $B_0$ = -0.35 mT. (**a**) Schematic of the lattice used as a reservoir. Blue readout lines indicate the row-by-row averaged magnetic state readout. (**b**) Inverse mean squared error 1/MSE as a function of the input magnetic field frequency and amplitude $B_1$. Higher (lower) values in yellow (purple) indicate better (worse) performance. The black contours correspond to 1/MSE ~ 30, highlighting regions where the reservoir performs better than the linear transformation of the original input signal. (**c**) Temporal view of the transformation. The upper panel shows the input sinewave signal (black dotted curve) and the target sawtooth signal (black dashed line). Shown in the lower three panels are the prediction with and without the reservoir (blue and red curve, respectively) for different input frequencies with the target sawtooth signal given for comparison (black dashed line).

The 1/MSE for a range of frequencies and amplitudes of the input magnetic fields $B_1$ is shown in Figure 6b. The black contour indicates a threshold of 1/MSE ~ 30, which can be achieved using only a linear transformation and without a reservoir. The best results (ii) are achieved at the input frequency of ~100 Hz. For other frequencies (e.g. at (i) 0.05 Hz and (iii) 100 kHz), the performance is noticeably worse and cannot be improved to the same level by altering the input field magnitude $B_1$ (Fig. 6b). Interestingly, the best performance is achieved at frequencies around the transition between the first ($m_{avg} > 0$) and the second ($m_{avg} \sim 0$, $q_{NN} \rightarrow 1$) ordering stages of the lattice (Panel 8 of Fig. 1b). A low-frequency input corresponds to the longer relaxation time (green-shaded region of Panel 8 in Fig. 1b) and allows the system to reach the same state with $m_{avg} \sim 0$ regardless of the input data. This results in a square rather than a sawtooth signal after the transformation (blue trace in panel (i) of Fig. 6c). For a high-frequency input (panel (iii) in Fig. 6c), the lattice cannot respond fast enough to the changes in the signal and cannot reproduce the rise or the sudden drop of the sawtooth. At an optimal input frequency (ii), the system achieves an MSE comparable to other magnetic reservoirs[67]. The performance can be further improved by increasing the size of the lattice or introducing a gradient in the effective anisotropy (see Supplementary Information 8).

In summary, we have demonstrated that the performance of the 2D Ising reservoir depends on how well the ordering dynamics matches the frequency of the input. This is characteristic for this type of computation and, therefore, the ease with which one can engineer the relaxation timescale in the 2D Ising lattices is highly advantageous. This avoids the otherwise necessary and computationally expensive preprocessing of the input data[63,66,68,71].

**Discussion**

We have shown with Monte Carlo simulations that the timescale of the relaxation dynamics in 2D Ising lattices can be tuned by varying the effective anisotropy and magnetostatic coupling, and that this can be exploited to achieve an enhanced performance of these lattices for reservoir computing. This information allowed us to fabricate experimental arrays of magnetostatically-coupled perpendicular nanomagnets with the desired properties by engineering the energy landscape through careful selection of materials and geometries.

We have shown that we can achieve a high degree of spontaneous ordering in arrays of out-of-plane nanomagnets. Nevertheless, achieving such a high level of ordering on an even shorter timescale as well as reaching $q_{NN}$ = 1 would be beneficial. Placing our experimental results on the ordering phase diagrams in Figure 2 reveals a way to do so. Indeed, Region IV with $q_{NN} \sim 1$ is located below the experimental points (circle, triangle and cross in Fig. 2) indicating sufficient $E_{MC}$ and excessive $E_{EA}$. Therefore, in order to achieve a perfect ordering at a given timescale, we need a method to tune the $E_{EA}$ with more precision. While we have shown that $E_{EA}$ can be reduced by increasing the magnetic layer thickness $t_{Co}$, going beyond the nominal sub-0.1 nanometre precision of $t_{Co}$ used in this work is challenging. Therefore, to control the perpendicular magnetic anisotropy in Co/Pt multilayers more precisely in the future, one can use thermal annealing[72–75]. Furthermore, laser annealing[76,77] would provide a means to locally modify $E_{EA}$ and create spatially complex designs.

In addition to making permanent modifications to the magnetic properties, $E_{EA}$ can be temporarily changed by applying current. The generated spin-orbit torque does not lift the degeneracy of "up" and "down" magnetic states in the out-of-plane easy axis geometry[78,79]. Therefore, on application of a current, the Joule heat and spin-orbit torque (requiring the lattice to be placed on an appropriate layer) lower the energy barrier to switching without interfering with the $E_{MC}$-driven ordering. Experimental confirmation of this is detailed in Supplementary Information 9. Using voltage and current to reduce $E_{EA}$ would provide a way to dynamically change the relaxation timescale of the lattice, opening the way to create systems with an adaptive temporal response, which is a cornerstone of information processing in the brain[80,81] and brain-like computing[82,83]. In addition, by engineering the current density distribution (as discussed in Supplementary Information 7) or by applying voltage to the individual nanomagnets[84], a more local electrical control could be achieved.

We have shown that by careful choice of the layer number and thicknesses in out-of-plane artificial spin ice, a magnetostatic coupling energy $E_{MC}$ of tens of $k_B T$ can be obtained. The ability to create systems with such high coupling provides an alternative, more energy efficient method to link the nanomagnets compared with the external electronic connections often used in probabilistic computing schemes[47,85] or for time multiplexing in reservoir computing[63,66]. For such applications, a further increase in $E_{MC}$ could be beneficial and can be accomplished by adding a soft magnetic layer with in-plane anisotropy underneath the out-of-plane lattice[28]. We tested this approach on square arrays with $D_{NM}$ = 140 nm and observed a significant increase in the spontaneous ordering (Supplementary Information 10). Another approach to increase $E_{MC}$ is to optimise the shape of the nanomagnets (Supplementary Information 11). Such control of the coupling by altering the shape is uniquely suited to out-of-plane nanomagnets since, for in-plane magnets, a change of shape tends to give a change in the distribution of the demagnetization field.

This freedom to control both $E_{EA}$ and $E_{MC}$ locally will provide a means to create lattices with novel emergent properties. In terms of fundamental science, this means that the thermodynamics can now be studied in a host of new lattice geometries to reveal new phases and phase transitions. In addition to creating thermally-active out-of-plane systems, it will now be possible to pattern mixed lattices incorporating both thermally-active in-plane and out-of-plane nanomagnets. These do not necessarily have an equivalent in bulk crystal systems and are likely to display unusual collective phenomena.

In terms of applications, these time-dependent artificial spin lattices of perpendicular nanomagnets offer an exciting platform for reservoir or probabilistic computing in square Ising-like and more complex lattices. The fact that these systems can be electrically interfaced, adapted to the input frequency and have short-term memory based on thermal activity, means that they have all the properties required for next generation computing.

Our work therefore opens new vistas across the fields of spintronics, neuromorphic computing and artificial spin ice, providing precise control of the cooperative behaviour and a flexibility in the design that can be finely tuned for different computing applications.

## Methods

### Sample fabrication

The films were deposited at room temperature onto high-resistivity Si wafers with a natural oxidation layer using DC magnetron sputtering. A base pressure of ≤1×10⁻⁶ Pa and Ar gas pressure of 3 mTorr were used for the sputtering. The deposited films were processed into devices with electron beam lithography and Ar ion milling. 50 nm-thick 2% hydrogen silsesquioxane was used as an electron beam resist to achieve high-resolution patterning of the nanomagnets. Poly(methyl methacrylate) 4% 950k was used as an electron beam resist for patterning the Hall bars and electrodes. For this, a milling current of 60 mA and angle of 5° were used.

### MFM Measurements

All MFM measurements were performed at room temperature. The samples were initialized by applying three 5-second-long pulses of a 0.9 T out-of-plane magnetic field, sufficient to saturate them and give a uniformly magnetized state on removing the magnetic field before the MFM measurements were carried out. Care was taken to make sure that there was no lateral magnetic field component or remanent magnetization associated with the magnetic field source that would affect the magnetic state of the array. All MFM scans were performed at zero field.

### Energy landscape calculations

The energy landscapes were calculated in Mathematica using a custom code. The profiles were calculated for the polar angle of magnetic moment of the nanomagnet going from 0° to 180° in 1° steps. To obtain the saturation magnetization of 1063 kA/m and interfacial anisotropy of 1.46 mJ/m$^2$ used for the energy landscape calculations, the thin films were measured using superconducting quantum interference device vibrating sample magnetometry (SQUID VSM). Details of the energy landscape calculations are given in Supplementary Information 4.

### "Hotspice" Monte Carlo simulations

The relaxation of the 2D Ising lattices and reservoir capability were simulated in Python using a custom "Hotspice" Monte Carlo code and employing a magnetic moment of 2.37×10⁻¹⁶ A m$^2$ for each spin and a temperature of 300 K. 600 input periods were used for the linear regression. 400 input periods were used to test the performance of the signal transformation. Details of the simulations are given in Supplementary Information 1.

### Acknowledgements

This work was funded by the EU FET-Open RIA project SpinENGINE (Grant No. 861618) and the Swiss National Science Foundation (Project No. 200020_200332).

### Author contributions

A.K, L.J.H. and A.H. conceived the project. A.K. fabricated the samples, designed and performed the experiments with support from A.H. and A.P. Analysis and interpretation of the data was done by J.M., A.K. and G.M.M. with support from all authors. J.M. and B.V.W. performed the Monte Carlo simulations and fitting of the experimental data. A.K. performed the Mathematica calculations and COMSOL simulations. A.K. and J.M. wrote the manuscript with input from all authors.

### Competing interests

The authors declare no competing interests.

SUPPLEMENTARY INFORMATION

Perpendicular-anisotropy artificial spin ice with spontaneous ordering: a platform for neuromorphic computing with flexible timescales

Aleksandr Kurenkov[1,2], Jonathan Maes[3], Aleksandra Pac[1,2], Gavin Martin Macauley[1,2], Bartel Van Waeyenberge[3], Aleš Hrabec[1,2] and Laura Jane Heyderman[1,2]

[1]Laboratory for Mesoscopic Systems, Department of Materials, ETH Zurich, 8093, Zurich, Switzerland
[2]Paul Scherrer Institute, Center for Neutron and Muon Sciences, Forschungsstrasse 111, 5232 Villigen, Switzerland
[3]DyNaMat, Department of Solid State Sciences, Ghent University, Belgium

## S1. "Hotspice" Monte-Carlo simulation details

The simulations were performed using the Hotspice[1] software package, which models the system as a lattice of dipolar-coupled Ising spins at a temperature $T$. The spins switch randomly as determined by the Néel-Arrhenius law $\tau_j(t) = \tau_0 \exp\left(\frac{\Delta E_j(t)}{k_B T}\right)$. The energy barrier to switching of an isolated spin $\Delta E = E_{EA}$. However, the interaction with other nanomagnets ($E_{MC}$) or an external magnetic field ($E_{Zeeman}$) affects the energy landscape and thus the effective barrier. In the limit of a zero-energy barrier, $\Delta E \to 0$, the switching time $\tau_j(t) \to \tau_0$. We choose $\tau_0 = 10^{-10}$ s, which is the timescale of the natural period of gyromagnetic precession of the magnetization about the easy axis of the nanomagnet[2].

The time evolution of the magnetic state of the lattice was evaluated stepwise, one spin switch at a time. Which spin switches next and how much later after the previous switching event it happens, is determined as follows. First, $\Delta E(t)$ is calculated for each nanomagnet individually, based on the lattice magnetization state. The corresponding average switching times $\tau_j(t)$ readily follow from the Néel-Arrhenius law. Then, for each magnet $j$, a random time $\Delta t_j$ is taken from an exponential distribution with mean value $\tau_j$. The nanomagnet with the smallest switching time $\Delta t = \min_j(\Delta t_j)$ will then switch if $\Delta t < t_{max}$, and the elapsed time $t$ increases by $\min(\Delta t, t_{max})$. The purpose of $t_{max}$ (default value is 1 second) is to avoid loss of accuracy for time-dependent external fields. For example, when a sinusoidal signal of frequency $f$ is applied to the lattice $t_{max} = 20/f$ ensures the simulation captures the waveform in enough detail.

## S2. The mechanism of the ordering slowdown after $m_{avg} \sim 0$ is reached

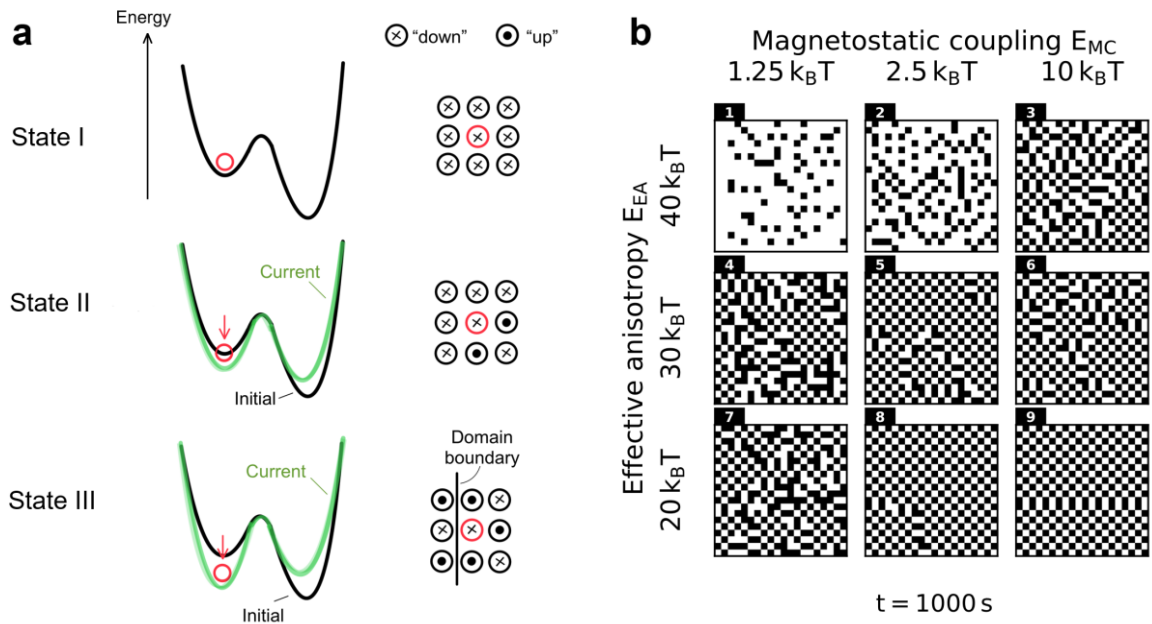

**Figure S2**. (**a**) Schematics of the energy landscape of the nanomagnet highlighted in red depending on the magnetic state of its neighbours. (**b**) Magnetic states of a 20×20 lattice at $t$ = 1000 s. White (black) contrast corresponds to the nanomagnets with "up" ("down") magnetic states.

The schematics in Figure S2a outline the energy-driven mechanism of the ordering slowdown after domain boundaries in the system have formed.

**State I:** the uniformly initialized state at $t$ = 0. It is easy for the central red spin to switch, as well as for its neighbours. These initial switching events start happening all over the system, eventually forming magnetically ordered areas (domains).

**State II:** assuming two of the neighbours of the red spin have switched, the red spin sinks deeper into its energy minimum. The stronger $E_{MC}$ is, the more pronounced this change in the energy minimum will be. It is now harder for the red nanomagnet to change its magnetic state.

**State III:** after some additional switching in the lattice, a domain boundary has formed, and the red spin is now a part of it. In order to reach perfect ordering, the red spin (or other similar spins in the domain boundary) would have to switch. However, it is now much more energetically expensive to switch than during the initial phase of individual switching events since the energy well is now deeper. This does not prevent the ordering process but significantly slows it down, resulting in the change of slope of $q_{NN}$ (the orange traces) in Figure 1 of the main text.

Figure S2b is the same as Figure 1c of the main text but for a larger 20×20 lattice. The extended lattice provides an additional overview of the formation and expansion of domains.

## S3. Dependence of $m_{avg}$ and $q_{NN}$ on $E_{MC}$, $E_{EA}$ and time following uniform initialization

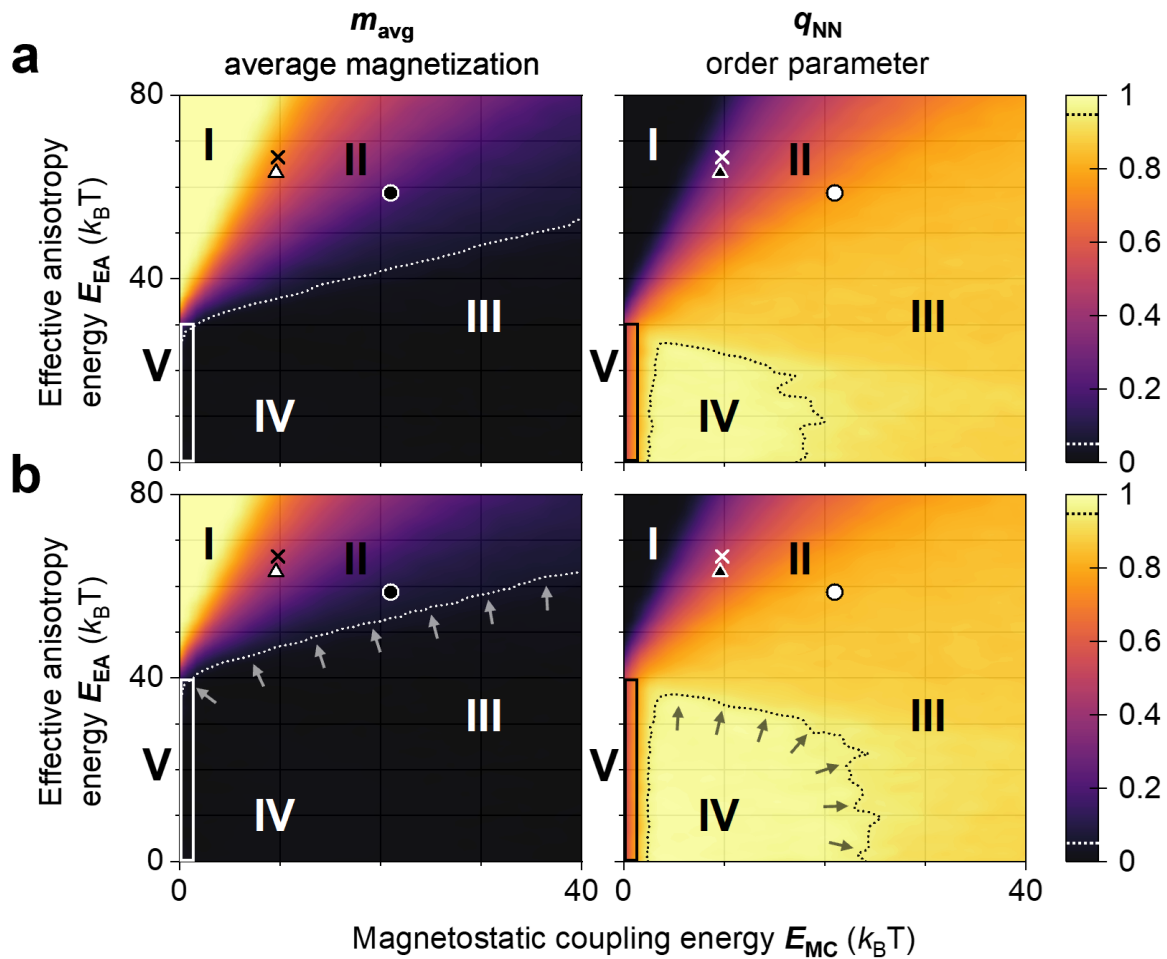

**Figure S3.** Phase diagrams of the average magnetization $m_{avg}$ and order parameter $q_{NN}$ as a function of the effective anisotropy $E_{EA}$ and magnetostatic coupling $E_{MC}$ (**a**) at $t \sim 1000$ s and (**b**) at $t \sim 7 \times 10^7$, which is ~27 months. The dotted lines are at values of 0.05 (white) and 0.95 (black). The changes in the position of the lines are highlighted by arrows. The labels for the five regions (I – V) and the experimental points (triangle, cross, circle) are the same as in Figure 2 of the main text. The vertical box showing Region V is the same in all panels.

Phase diagrams of the average magnetization $m_{avg}$ and order parameter $q_{NN}$ at $t \sim 1000$ s (Fig. S3a) and $t \sim 7 \times 10^7$ (Fig. S3b). Note the shifting of the borders of Regions III, IV and V as time increases. The border of Region V in Figure S3b is now located at $E_{EA} \sim 41$ $k_BT$, because nanomagnets with this effective anisotropy have an average switching time of $\sim 7 \times 10^7$ s. Note that these phase diagrams were calculated assuming that there is no exchange coupling between the nanomagnets, so $J = 0$. Therefore, since the $S_{ASI} = 20$ nm 'circle' experimental point is for the lattice with $J \sim 6.5$ mT, its location is approximate.

**S4. Calculation of energy landscape of coupled Co/Pt nanomagnets**

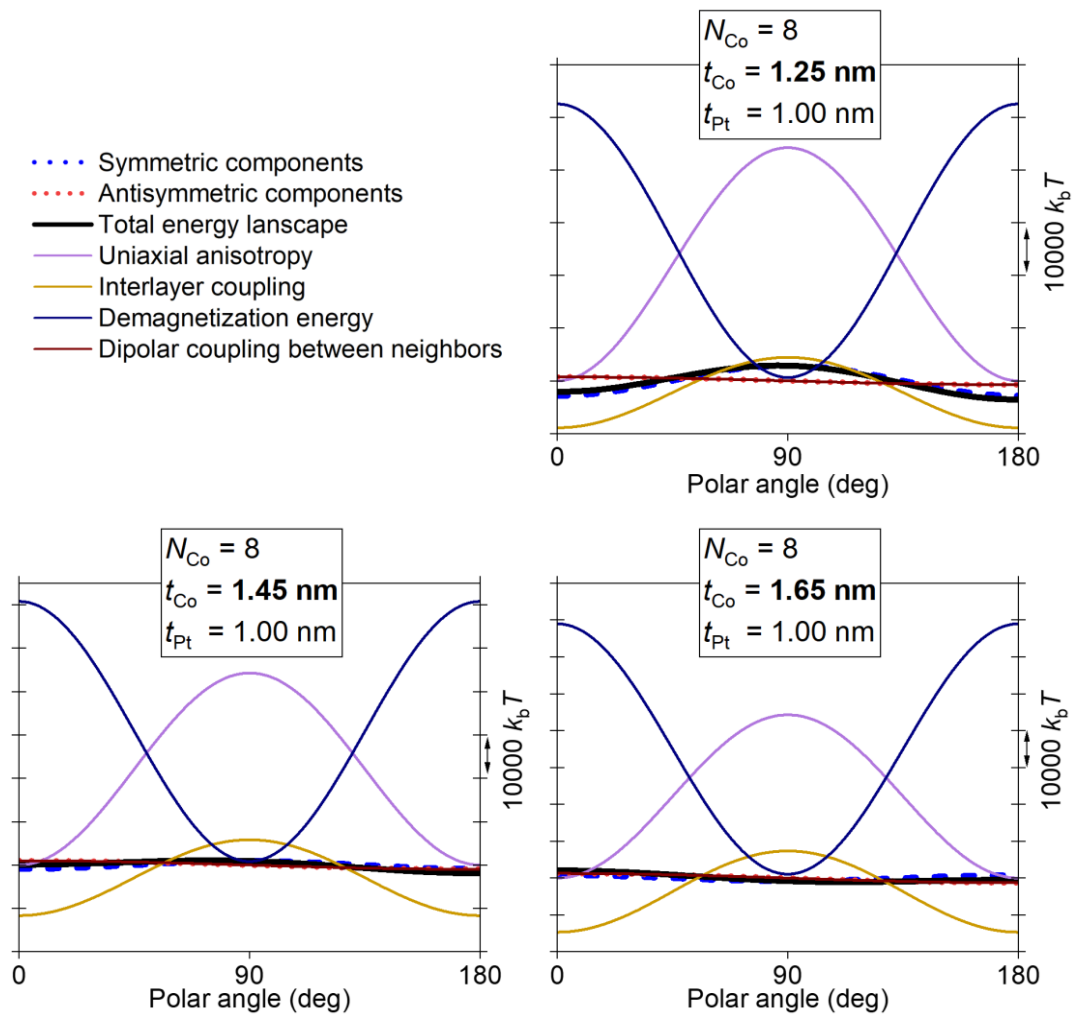

**Figure S4**. Energy landscapes for the same cases shown in Figure 3(e)-(g), but now with all four energy terms.

The energy landscapes were calculated in Mathematica for polar angles from 0° to 180° in 1° steps. Four energy terms were considered associated with (i) the uniaxial anisotropy, (ii) the demagnetization, (iii) the interlayer magnetostatic interaction within the nanomagnet and (iv) the dipolar coupling to the neighbouring nanomagnet. The first three terms contribute to effective anisotropy $E_{EA}$ and the last one is the only contributor to the magnetostatic coupling term $E_{MC}$. A uniaxial anisotropy of 1.46 mJ/m² was determined from vibrating sample magnetometry measurements of the multilayer films with $N_{Co} = 3$, $t_{Co} = 1.2$ nm and $t_{Pt} = 0.4$ nm. For the other three terms, a saturation magnetization of 1063 kA/m, derived from the same measurements, was included. The dipolar coupling between each pair of nanomagnets was calculated by direct integration. The interaction between the layers within one nanomagnet was calculated following Dmytriiev *et al*. (2010)[3]. The demagnetizing energy was calculated using the demagnetization tensor from Joseph (1966)[4].

Shown in Figure S4 are the energy landscapes of Figure 3(e)-(g) with all four energy components plotted along with $E_{EA}$, $E_{MC}$ and ($E_{EA}+E_{MC}$). From the relative magnitudes of the four energy components compared to the final landscape, one can appreciate the precision required in balancing the demagnetizing energy with effective anisotropy to achieve a desired landscape. Another interesting observation is the non-negligible role of the interlayer coupling within a nanomagnet (yellow trace).

Note that the energies obtained in these energy landscape calculations are not the ones used in the "Hotspice" Monte-Carlo simulations. This is because the energy landscape simulations overestimate the energy as they assume a simplified reversal model of coherent rotation of the magnetization, rather than domain wall nucleation and propagation. Nevertheless, this simple model gives important information of the dependence of the energy landscape on the lattice parameters. For the Monte Carlo simulations, we cover a parameter space that is relevant to our experimental systems.

**S5. Fitting of the experimental MFM data for $D_{NM}$ = 170 nm with Monte Carlo simulations**

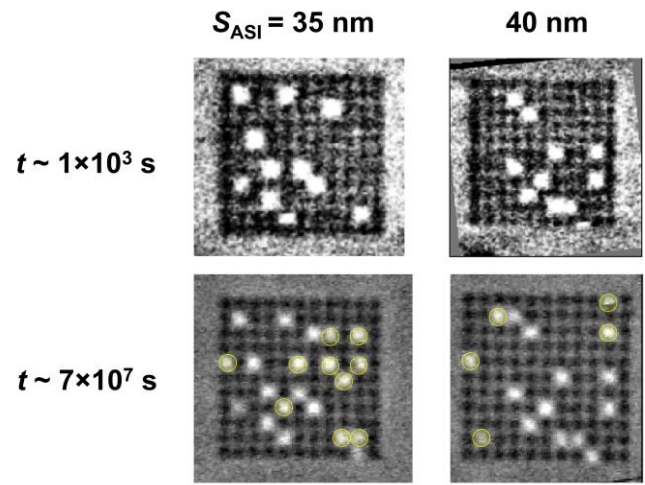

**Figure S5.1**. MFM images of lattices with $D_{NM}$ = 170 nm, $t_{Co}$ = 1.45 and $S_{ASI}$ = 35 and 40 nm at $t$ = 1000 s and $t$ = 7×10$^7$ s. The magnets that have switched between these observations are highlighted with yellow circles.

In Figure S5.1, MFM images are given for the lattices with $D_{NM}$ = 170 nm and nanomagnet separations, $S_{ASI}$, of 35 and 40 nm. The measurements were performed in the same way and at the same time as the images in Figure 4 of the main text. This data was used to determine the corresponding $m_{avg}$ and $q_{NN}$ in Figure 4f, as well as $q_{2NN}$ and $q_{3NN}$.

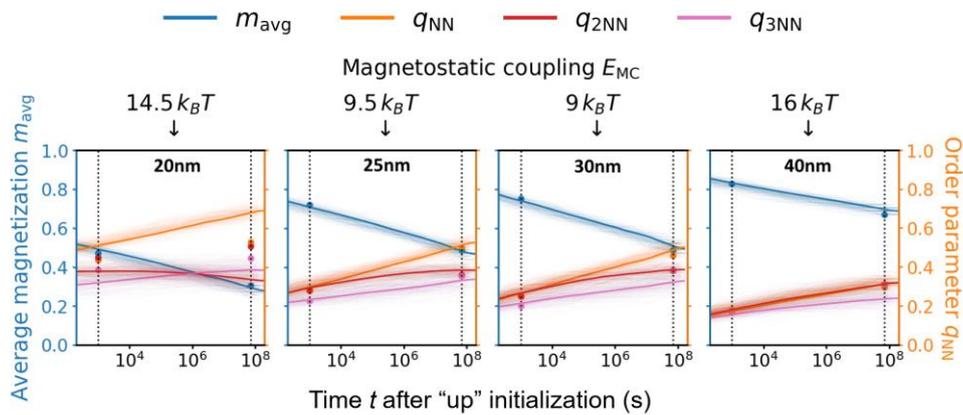

**Figure S5.2**. Evolution with time of the average magnetization $m_{avg}$ and order parameters $q_{NN}$, $q_{2NN}$, $q_{3NN}$. The fine lines are calculated using Monte Carlo simulations assuming point dipoles and no exchange coupling. The bold lines are their averages. The points are experimental data. $E_{MC}$ obtained from the fitting are shown in the figure. $E_{EA}$ obtained from the fitting is ~65.0 $k_B$T.

We then performed a fitting of the Monte Carlo simulated curves to the experimentally determined values of $m_{avg}$, $q_{NN}$, $q_{2NN}$ and $q_{3NN}$ at $t$ ~ 1000 s and t ~ 7×10$^7$ s. The simulation was carried out in exactly the same way as to calculate relaxation dynamics graphs shown in Figure 1b of the main text, and we used $E_{EA}$ and $E_{MC}$ as fitting variables. This fitting was performed using a weighted least-squares method, where the weights for the variables $m_{avg}$, $q_{NN}$, $q_{2NN}$ and $q_{3NN}$ at each time point $t$ were determined based on the variability observed in the Monte Carlo simulations. Specifically, each variable was assigned a weight inversely proportional to its standard deviation at that time. The smaller the standard deviation, the higher the weight, implying a higher fit quality. The results are shown in Figure S5.2.

As can be seen in Figure S5.2, a better fit for $S_{ASI}$ = 25 and 30 nm could be achieved than for $S_{ASI}$ = 20 nm. As explained in the main text, this could be due to a residual Co layer between the nanomagnets that results in some ferromagnetic exchange coupling between them. To account for this, we introduced exchange coupling $J$ into the fitting procedure. This $J$ gives a reduction in the nearest-neighbour dipolar coupling. We also accounted for the finite size of the nanomagnets by adding a $1/r^5$ term to $D_{ij}$ (Eq. 2 of the main text) as described in Reference [5]. This resulted in a significantly better fit for the $S_{ASI}$ = 20 nm case, as shown in Figure S5.3a. As expected, the fitting procedure returned a non-zero exchange coupling of ~ 6.5 $k_B$T for the lattice with $S_{ASI}$ = 20 nm and near-zero exchange coupling for lattices with $S_{ASI}$ = 25 and 30 nm.

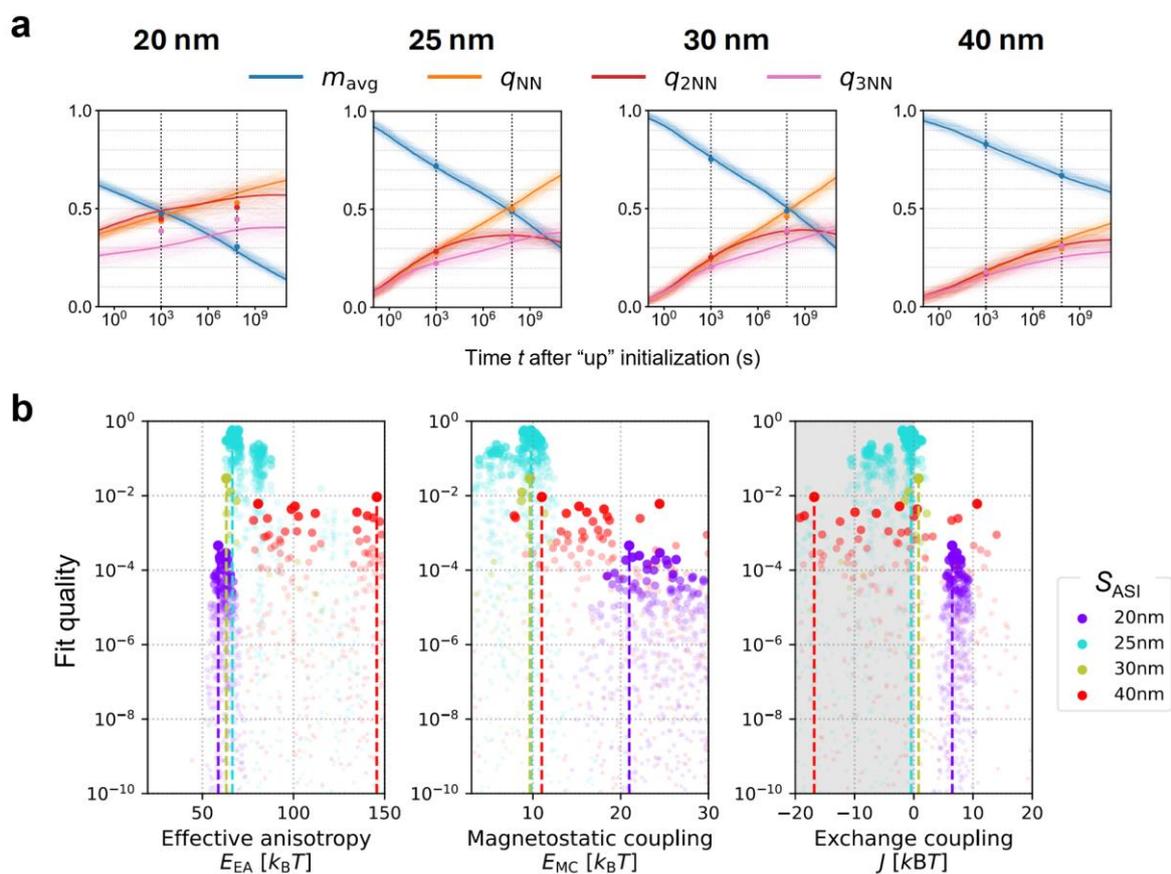

**Figure S5.3**. Fitting for the experimental data that accounts for non-zero exchange coupling $J$. Panel (**a**) shows the fitted results; the lines and points are as in Figure S5.2. Panel (**b**) shows how well the Monte Carlo simulation fits the experimental data at a given $E_{EA}$, $E_{MC}$ and $J$ value (left, middle and right panel, respectively). A larger y-axis value (fit quality) indicates a better fit. For example, for $S_{ASI}$ = 20 nm (purple points) and $E_{EA}$ (left plot), the highest y-values are concentrated around $E_{EA}$ = 60 $k_B$T, indicating that the best fit for this lattice can be achieved at $E_{EA}$ of ~60 $k_B$T. Greyed-out area represents antiferromagnetic coupling (negative $J$), which is considered non-physical in our system.

Despite the introduction of $J$, we could not obtain a good fit for the $S_{ASI}$ = 40 nm. Figure S5.3b shows how well the experimental data can fitted at a given $E_{EA}$, $E_{MC}$ and $J$ (left, middle and right panel, respectively). In particular, each data point in Figure S5.3b corresponds to a fit of the experimental data to the Monte Carlo simulation (fit quality), and the larger the y-axis value, the better the fit. Whereas for $S_{ASI}$ = 20, 25 and 30 nm, the best fits are mainly concentrated in a rather narrow range of $E_{EA}$, $E_{MC}$ and $J$ (cyan, purple and green points), this is not the case for $S_{ASI}$ = 40 nm (red points). Indeed, for $S_{ASI}$ = 40 nm, fits of equal quality could be obtained for several $E_{EA}$, $E_{MC}$ and $J$, with many of the red points taking similar y-axis values.

## S6. Experimental electrical readout of out-of-plane artificial spin ice

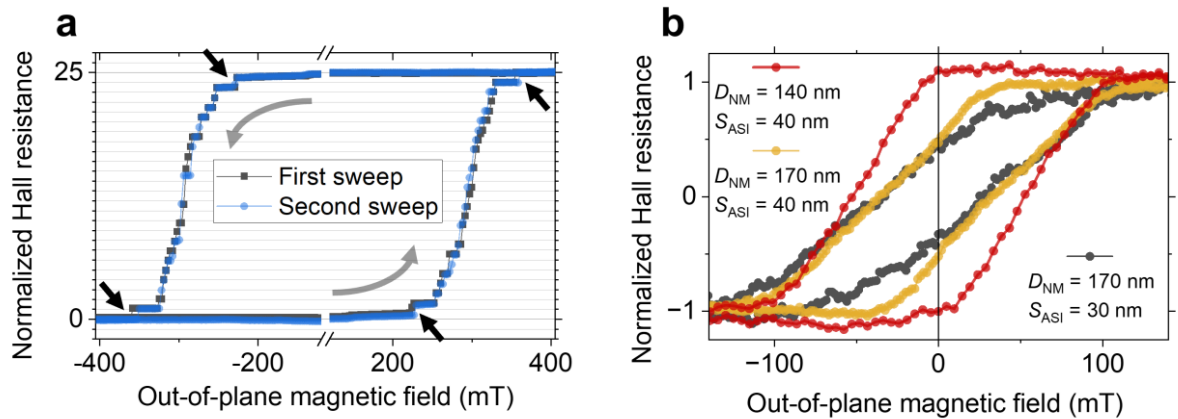

**Figure S6 | Electrical readout of square lattices of perpendicular nanomagnets.** The lattices were placed on Hall bars. The magnetic field was swept while probing anomalous Hall voltage. The grey arrows indicate the direction of the magnetic field sweep. (**a**) Hall resistance measurements on varying the out-of-plane magnetic field for a lattice of 5×5 magnets with $t_{Co}$ = 1.2 nm and $D_{NM}$ = 800 nm that did not show spontaneous switching at the timescale of the experiment. The readout was normalized from 0 to 25 to highlight the 25 binary switching events associated with the 25 individual nanomagnets. The black arrows point to the first and last switching event when increasing and decreasing the magnetic field. (**b**) Hall resistance measurements on varying the out-of-plane magnetic field for a lattice of 11×11 magnets with different nanomagnet diameters $D_{NM}$ and separations $S_{ASI}$. A Keithley 6221 current source and 2182 nanovoltmeter were used for electrical measurements.

Electrical interfacing is essential for practical applications as well as for fundamental investigations of the magnetization dynamics in artificial spin ice[6], providing a means to probe the dynamics in systems of any size in a time-resolved fashion[7]. However, electrical interfacing is rarely used to probe nanomagnets with in-plane easy axis since this cannot be carried out without the help of additional layers[6], second-harmonic readout[8], external probing of the magnetic field[8] or performing measurements at temperatures below 50 K[9].

In contrast to in-plane systems, the out-of-plane anisotropy of the nanomagnets in our artificial spin ice allows us to directly perform electrical readout as they are. To demonstrate this, we placed several different lattices on Hall bar electrodes and applied a small probing current of ~4×10^9 A/m². The generated transverse anomalous Hall voltage was then divided by the current to obtain the Hall resistance $R_{Hall}$, which represents the magnetic state of the system and is approximately equivalent to $m_{avg}$ (the difference between $R_{Hall}$ and $m_{avg}$ is discussed in Supplementary Information 7).

We first measured a 5×5 square lattice of circular nanomagnets with $D_{NM}$ = 800 nm patterned from a (Co [1.2] / Pt [0.2])$_2$ / Co [1.2] multilayer, which did not show spontaneous ordering in the MFM images. We swept the out-of-plane magnetic field while probing $R_{Hall}$ (Fig. S6a). The readout was normalized from 0 to 25 in order to highlight the 25 binary switching events associated with each of the 25 nanomagnets. The first (last) switching event in each of the branches happens noticeably earlier (later) than the rest (shown by black arrows in Fig. S6a). This may be an indication of the magnetostatic coupling between the nanomagnets as the first (last) switching event is the most (least) energetically favourable due to the magnetic configuration of the neighbouring nanomagnets. However, this system is not thermally active at the measurement timescale of tens of seconds. The absence of thermal activity is apparent from the same value of $R_{Hall}$ (and thus $m_{avg}$) for the remanent and saturated state so that, once the field is removed, the lattice remains frozen.

We then performed the same measurement on 11×11 square lattices with $D_{NM}$ = 140 and 170 nm, $S_{ASI}$ = 30 and 40 nm, and a multilayer stack of (Co [1.45] / Pt [0.8])$_6$ / Co [1.45] (Fig. S6b). We know that these lattices show spontaneous switching at the timescale of seconds (see "Control of relaxation timescales in experimental dipolar-coupled 2D Ising systems" section of the main text). We also know that these nanomagnets are single-domain and have a perpendicular easy axis from the MFM data (Fig. 4a and Fig. 5d of the main text). A remanent $R_{Hall}$ (~$m_{avg}$) that is lower than saturation is thus only possible due to spontaneous antiparallel magnetic ordering of the nanomagnets, and a smaller remanence indicates that more nanomagnets have switched. For the sample with $D_{NM}$ = 140 nm and $S_{ASI}$ = 40 nm (red loop in Fig. S6b), the lattice remains in the saturated state after the field is removed, whereas increase of $D_{NM}$ to 170 nm results in a smaller remanent $R_{Hall}$ ($m_{avg}$) and onset of spontaneous switching (yellow loop in Fig. S6b). Decreasing $S_{ASI}$ from 40 nm to 30 nm for $D_{NM}$ = 170 nm further lowers the remanent $R_{Hall}$ (~$m_{avg}$) (dark grey curve in Fig. S6b).

Therefore, as we go from lattices with weaker magnetostatic coupling between the nanomagnets to those with stronger coupling ($D_{NM}/S_{ASI}$ = 140 /40 nm → $D_{NM}/S_{ASI}$ = 170/40 nm → $D_{NM}/S_{ASI}$ = 170/30 nm), we observe a decrease in both $m_{avg}$ (measured by MFM) and remanent $R_{Hall}$ (measured electrically). The trends observed in the electrical measurements are thus consistent with the MFM results, providing a justification for using electrical reservoir readout in the simulations of a 2D Ising reservoir in the "Tuneable-frequency reservoir computing with 2D Ising systems" section of the main text.

**S7. Simulation of local electrical readout with multiple electrodes**

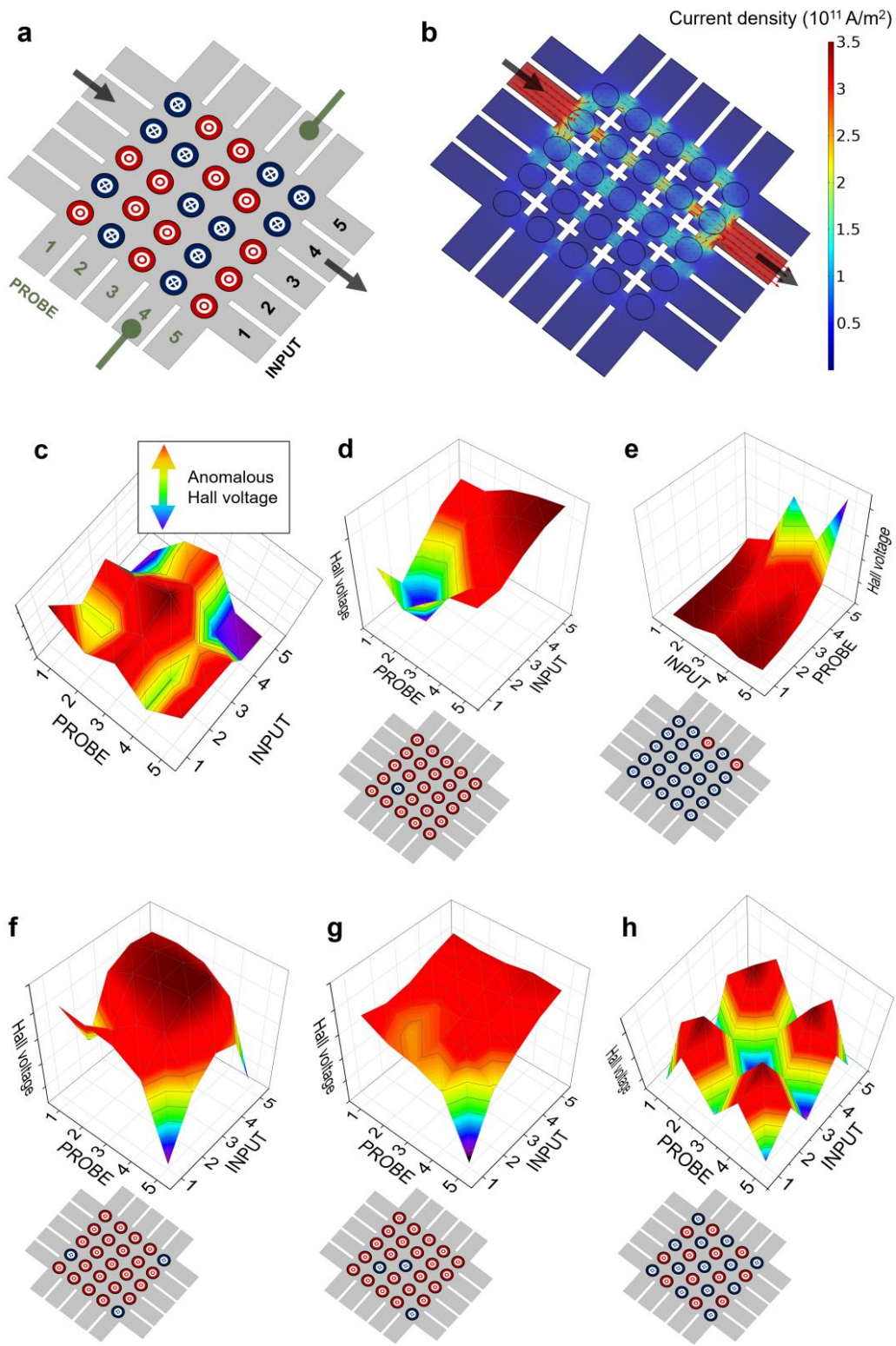

**Figure S7**. Local readout with planar electrodes. (**a**) Schematic of the electrodes and the readout scheme. Red (blue) circles represent nanomagnets with "up" ("down") magnetization. (**b**) An example of a non-uniform current distribution in the system. (**c**) Readout pattern produced by the magnetic state shown in (a). (**d**)-(**h**) Readout patterns corresponding to different magnetic states indicated to the bottom left of the plots. The increment of the vertical axis is 2 mV in all cases.

Inhomogeneities in the current density distribution in the Hall bar used to measure the magnetic state of the 2D Ising systems may result in different currents flowing through the different nanomagnets, resulting in unequal contributions to $R_{Hall}$. This may lead to a distortion in the one-to-one correspondence between $R_{Hall}$ and $m_{avg}$. This effect is negligible for lattices of limited size placed on a symmetric Hall bar, as can be seen from the 25 rather homogenous steps in $R_{Hall}$, corresponding to single nanomagnets switching, in Figure S6a. However, the inhomogeneity in the current density can be deliberately increased by adding more electrodes to the system, and this can be used to extract individual states of the nanomagnets and not just $m_{avg}$. To demonstrate this, we performed COMSOL simulations of a 5x5 lattice with 5 electrodes on each side, organized into 5 pairs of "input" electrodes for applying current and 5 pairs of "probe" electrodes for measuring the resultant anomalous Hall voltage (Fig. S7a). Changes in current distribution, which depend on which electrodes are used, result in the probing of different areas of the lattice. For example, applying a voltage to the 3$^{rd}$ pair of input electrodes results in the current density distribution shown in Figure S7b.

We then applied a test sequence, which consisted of applying a probing current to a pair of input electrodes while reading out Hall voltage from one of the pairs of transverse probe electrodes. This was repeated for all the combinations of input and probe electrode pairs, giving a matrix of 25 measurements. A representation of this readout is shown in Figure S7c, where x and y coordinates are input and probe pair number, and the z (vertical) coordinate is the corresponding anomalous Hall voltage readout. The readout in Figure S7c corresponds to the magnetic state of Figure S7a. Nanomagnets that are magnetized "up" ("down") produce a positive (negative) contribution to the readout, as one can see by comparing Figures S7a and S7c, particularly in the corners of the lattice. However, there is no one-to-one correspondence between the magnetic state and the electrical readout because every nanomagnet contributes to the Hall voltage for each of the 25 input-probe electrode combinations. Nevertheless, the magnetic state of the lattice is clear from the readout pattern in some simple cases such as those shown in Figures S7d, e and f, in which peaks or troughs correspond to the few nanomagnets magnetized in the opposite direction to the majority of the lattice. The more complex magnetic states are difficult to understand from the readout pattern (e.g. in Fig. S7g; in Fig. S7h some peaks correspond to nanomagnets with "down" magnetization due to the contribution of the surrounding nanomagnets).

To understand the Hall voltage readout for the general case of arbitrary magnetic states, we simulated readout patterns where only one single nanomagnet has a magnetization orientation that differs from the others, which we call "special readout patterns". There are 25 such special readout patterns, and one example is shown in Fig. S7d. For the general case, any Hall voltage readout, such as those illustrated in Fig. S7c to S7h, can be constructed by summing these 25 special readout patterns weighted by a +1 or -1 coefficient, which reflects the orientation of the the single nanomagnet with differing magnetization orientation in the array. There is not always a direct correlation between the peaks and troughs of the general readout patterns, and the "up" or "down" states of nanomagnets, as can be seen in Fig. S7h. However, for certain nanomagnets, the sign of the Hall voltage is always correlated with their magnetic state. This is true for the nanomagnets near the input and probe electrodes, which contribute most significantly to the readout and thus do not get masked by contributions from their neighbours. Identifying such cases allows us to remove their contributions with the correct signs from the general readout, and then to proceed with the next-largest contributors. We describe this in more detail as follows:

- The nanomagnets in the corners of the lattice provide the largest contributions to the readout, overriding any signal from the rest of the system. The sign of the readout for the combination of inputs 1 and 5 with probes 1 and 5 thus corresponds to the magnetic states of these nanomagnets. Using this, we determine their magnetic states ("up" or "down") and subtract these special readouts from the initial readout pattern with the corresponding (+1 or -1) signs.

- Once the contributions of the nanomagnets in the corners have been subtracted, the next-largest nanomagnet contributors are the sets of three nanomagnets between the pairs of corner nanomagnets (e.g., probe 1 in combination with inputs 2, 3, 4, or input 1 in combination with probes 2, 3, 4) since they are the closest to the electrodes among the remaining contributors. The magnetic states of these nanomagnets can thus be determined from the sign of the corresponding readout and then their corresponding special readouts can be subtracted accordingly.

- After these contributions have been subtracted, the eight nanomagnets forming a square around the central magnet, followed by the central one, can be read out and subtracted.

Peeling off the readout in this way provides a route to retrieve all of the states of the 25 individual nanomagnets from the electrical readout of any magnetic state. This means that any parameter of the system, such as $q_{NN}$, $q_{2NN}$ and $q_{3NN}$, can be determined using the planar electrodes. The temporal resolution of such measurements is limited by the time necessary to perform the 25 measurements with sufficient signal-to-noise ratio. Therefore, while $m_{avg}$ can be tracked with nanosecond resolution, measurements of $q_{NN}$, $q_{2NN}$ and $q_{3NN}$ may be slower.

From a computational perspective, these results are important because they show that, even if individual magnetic states are not read out directly (e.g., when using a crossbar), they can still be retrieved by performing a set of linear operations. In the context of reservoir computing, this means that the use of this readout approach with multiple electrodes can be computationally equivalent to knowing the exact magnetic configuration of the system.

Finally, it is worth mentioning that the maximum system size, to which this local electrical readout method with planar electrodes can be applied to determine the magnetic state, may be limited by the lower signal from central areas that are far from any electrodes. Maximizing the current density, for example, by adding cutouts in the electrode as shown in Fig. S7b, may be helpful in alleviating this problem. In addition, using machine learning instead of the sequential "peeling off" of the readout discussed here may also be helpful in extending this method to lattices of larger sizes or more complex geometries.

## S8. Transformation improvements and Mackey-Glass prediction

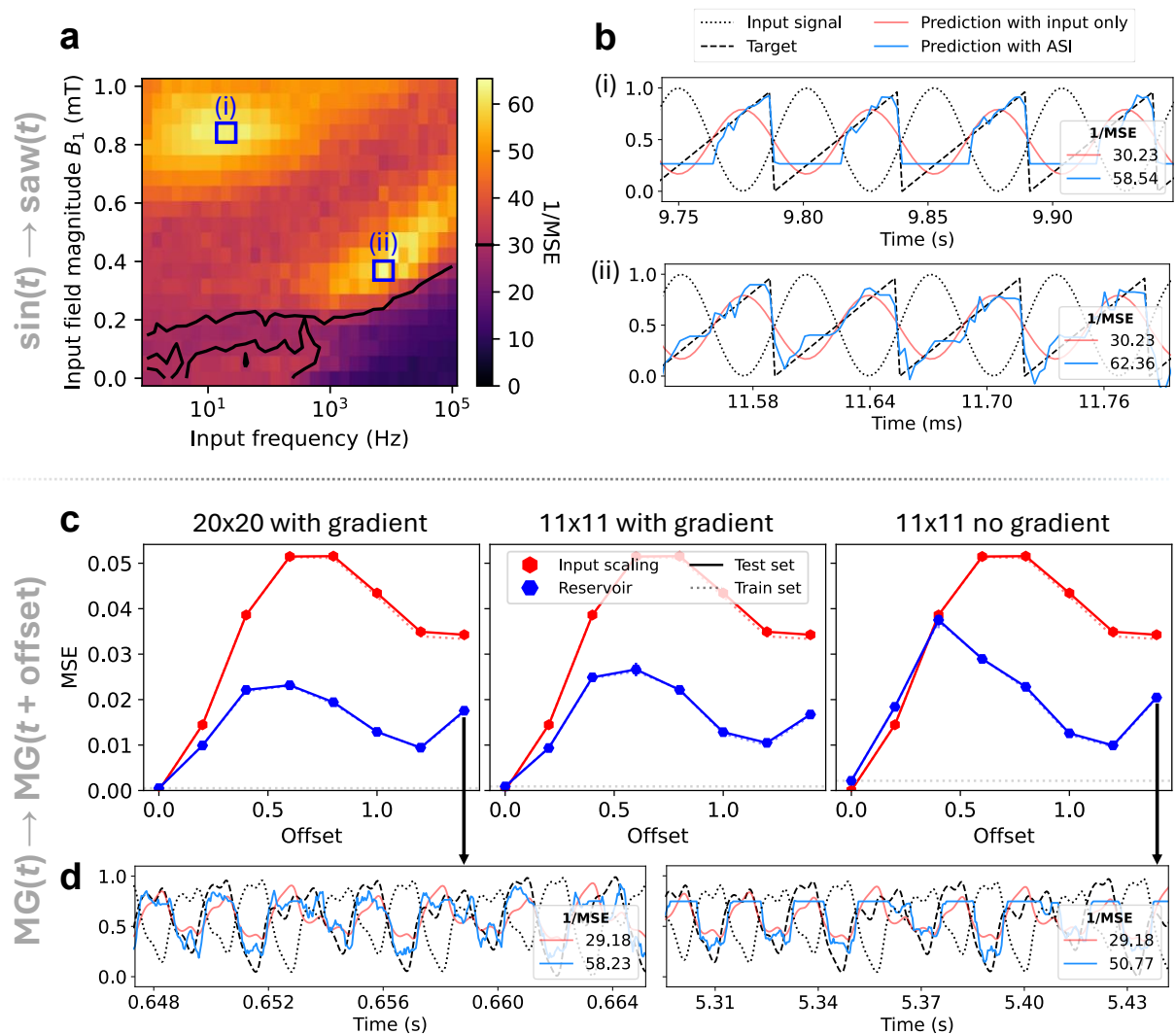

**Figure S8.** Signal transformation of different sized lattices with and without a gradient in $E_{EA}$ and magnetic moment. The parameters $E_{MC} = 2.5k_BT$ and $E_{EA} = 20k_BT \pm 5\%$ are employed for all figure parts. (**a**) Inverse MSE for a sine wave-to-sawtooth transformation for a 20×20 lattice with a 10% gradient in $E_{EA}$ and magnetization, as a function of input field frequency and maximum input field $B_1$. The minimum input field $B_0$ = -0.2 mT. (**b**) Temporal trace of the sine wave-to-sawtooth transformation for the two hotspots in (a). The input signal is given by black dotted lines and the target is given by black dashes. The prediction with (without) the reservoir is given by blue (red) lines. (**c**) Performance for a Mackey-Glass prediction given by MSE as a function of desired future time 'offset', for two system sizes with and without a gradient. (**d**) Temporal traces of the Mackey-Glass prediction with offset 1.4, both for the 20×20 lattice with a gradient and for the 11×11 lattice without a gradient.

The sine wave-to-sawtooth transformation presented in the main text was simulated for an 11×11 lattice of spins that is just like the experimental system of nanomagnets. However, several modifications to the lattice can improve performance. Firstly, enlarging the system is beneficial because this reduces thermal noise in the output, as each readout value becomes the average of more magnets. Secondly, introducing a gradient in effective anisotropy $E_{EA}$ and magnetic moment of the spins to obtain a greater variation in the readout values will also yield a better transformation. Here, we implement a gradient where $E_{EA}$ and magnetic moment of the spins gradually gets larger going from one side of the lattice to the other. This gradient in $E_{EA}$ means that magnets on one side of the lattice relax faster than on the other side, providing several different short-term memory timescales. A "gradient of *10%*" means that the $E_{EA}$ on the right (left) side of the lattice is *10%* higher (lower) than the average. The exchange coupling $J$ between the nanomagnets is set to 0.

The effect of extending the lattice and adding a gradient in $E_{EA}$ and magnetic moment is presented in Figure S8. Figures S8a and S8b concern a sine wave-to-sawtooth transformation (as in the main text), but for a larger 20×20 system with a gradient of 10%. Note that, while a low MSE (or high 1/MSE) is often indicative of a good transformation, this is not necessarily the case in noisy systems where the noise imposes a lower bound to the MSE. For example, in Figure S8a, there exist two distinct combinations of input magnitude and frequency that both give a good MSE (blue boxes), yet the temporal traces (Fig. S8b) reveal that the higher-frequency input (7.5 kHz / 0.37 mT) yields a much better sawtooth than the lower-frequency input (20 Hz / 0.84 mT). For the low-frequency input (upper panel), the magnetic state saturates at large amplitudes of the sine input, so that the low peak in the sawtooth is cut off. Nevertheless, the prediction follows the steep drop of the sawtooth better than with the high-frequency input (due to a finite relaxation time), thus giving similar MSE. This illustrates that, not only MSE should be considered as a metric for a high-performance reservoir, but that the frequency adjustment is also key for obtaining a correspondence in shape between the reservoir-based prediction and the target.

In Figures S8c and S8d, we explicitly compare the results from large and small systems, with and without a gradient in $E_{EA}$ and magnetic moment, but now for a time series prediction task. We use the standard task of a chaotic Mackey-Glass (MG) oscillator[10] whose states must be predicted at a given amount of time in the future, which is referred to as the 'offset'. The following parameters of MG equation were used: $\tau$ = 23.0, $\beta$ = 0.2, $\gamma$ = 0.1, n = 10.0 (Ref. [11]). The MSE of the various lattices as a function of MG offset, i.e., how far into the future should be predicted, is shown in Figure S8c. The datapoints show the best MSE among all lattices of a given class for a given offset. For instance, for offset = 1 of the panel "20x20 with gradient" in Fig. S8c, we tested the prediction task on 20×20 lattices with various gradient strengths while applying inputs with different frequencies $f$ and encoding fields $B_0$ and $B_1$, and then selected the best MSE among the results. The effect of adding a gradient in both $E_{EA}$ and the magnetic moment, and enlarging the system turn out to be significant. This improvement can be seen when comparing the temporal views in Figure S8d, where the larger system with a gradient (lefthand panel) produces a closer prediction than the smaller system without a gradient (righthand panel). For example, for the smaller system, it can be seen that, for the prediction with artificial spin ice (blue curve), the top of the peaks is cut off. This is not the case for the larger system. These results are similar to the performance of other magnetic reservoirs[12–14].

## S9. Experimental lowering of the switching energy with current-induced Joule heating and spin-orbit torques

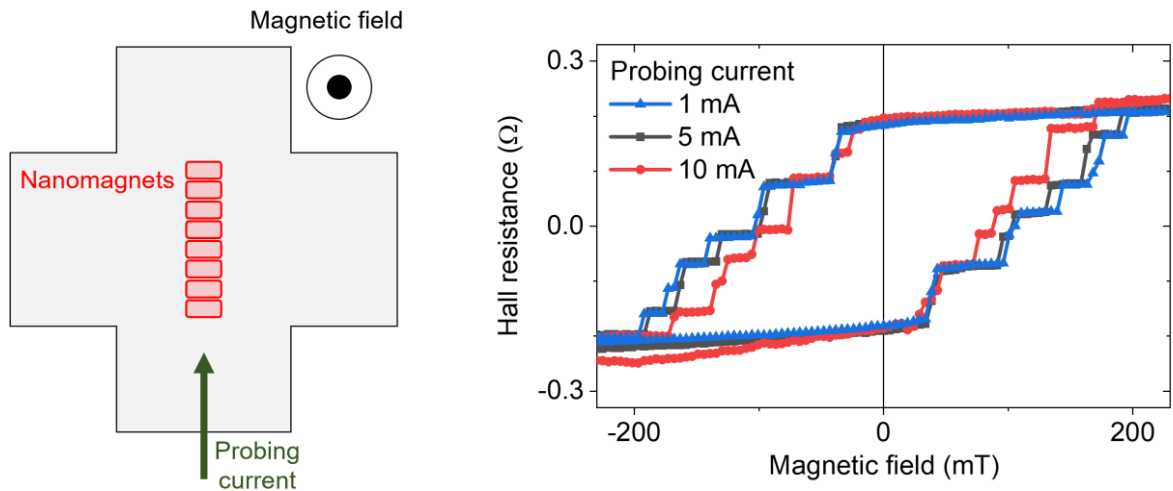

**Figure S9**. Hall resistance versus out-of-plane magnetic field for different probing currents.

Applying Joule heating and spin current to the lattice of nanomagnets with perpendicular magnetic anisotropy can lower the energy barrier to switching of the nanomagnets. Here we look at the effect of Joule heating and spin current on a 1D chain of 8 nanomagnets, similar to that shown in Fig. S11, but which does not show spontaneous ordering at room temperature at the experimental timescale. The lattice is placed on a Ta/Pt Hall bar. The current is used for probing and simultaneously acts as a source of Joule heating and spin-orbit torque, and we find that an increase in current leads to switching of the nanomagnets at smaller fields (Fig. S9). Since neither Joule heating nor spin-orbit torques break the symmetry between the "up" and "down" states, the energy barrier is modified equally for all nanomagnets, and the order in which the nanomagnets switch is governed by the magnetostatic coupling. Therefore, we imagine that this approach could be suitable for annealing artificial spin ices with perpendicular anisotropy, without the need to apply a magnetic field or heat.

## S10. Magnetostatic coupling enhancement with a Permalloy underlayer

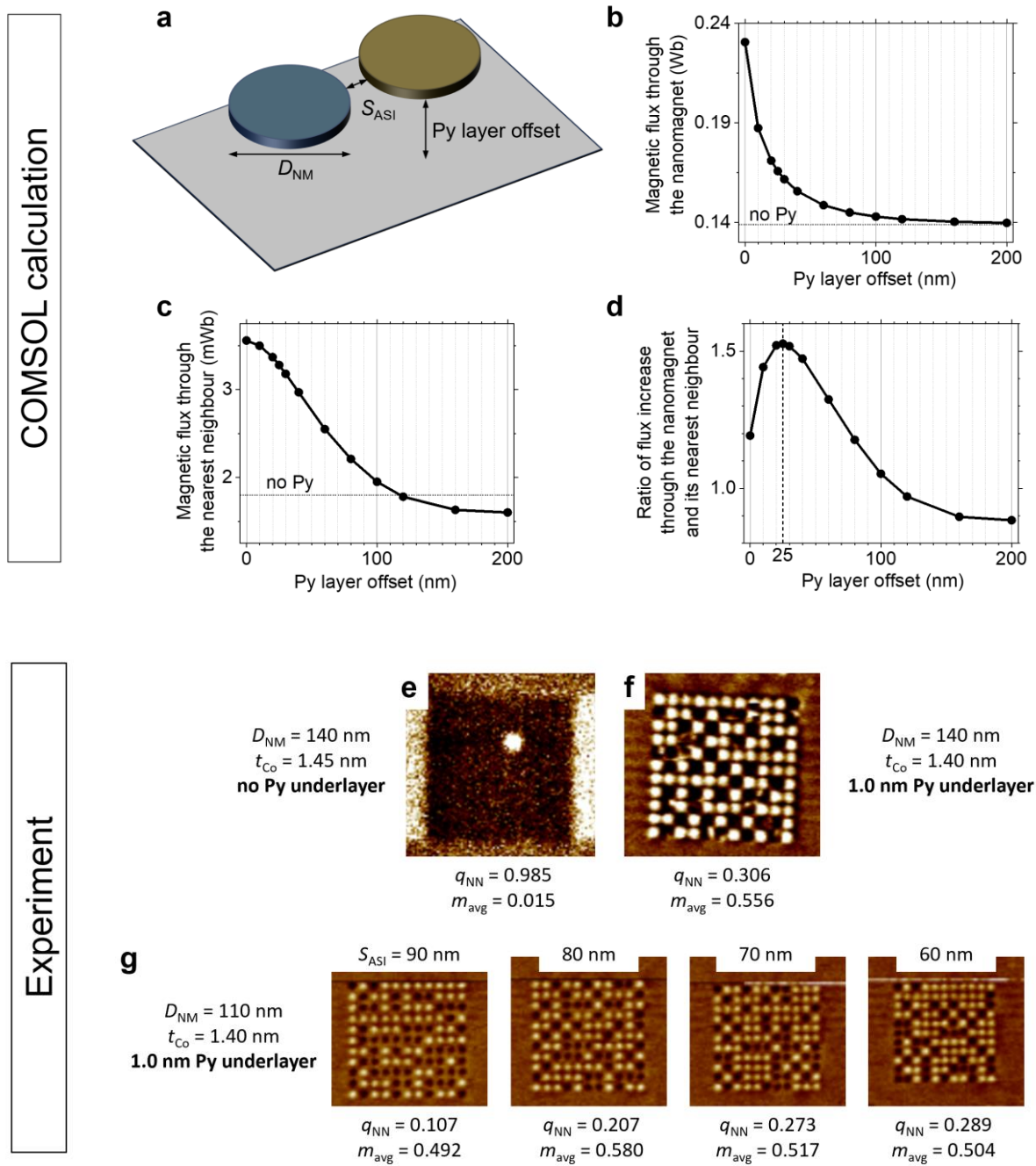

**Figure S10**. Enhancement of intermagnet coupling and spontaneous ordering through mediation of the magnetic flux with a permalloy underlayer. All images were taken at $t \sim 1000$ s. (**a**) Schematic of the simulation. (**b**) Change of magnetic flux through the nanomagnet itself with the distance to the Py underlayer. (**c**) Change of magnetic flux through the neighbouring nanomagnet with the distance to the Py underlayer. (**d**) Ratio between (c) and (b). (**e**) and (**f**) Spontaneous switching in 11×11 lattices in similar multilayer stacks without and with Py underlayer. The Py offset is 25 nm. (**g**) Spontaneous switching in an 11×11 lattice for nanomagnet diameter $D_{NM}$ = 110 nm and different separations $S_{ASI}$. The distance to the Py layer is 25 nm.

It is known that a Permalloy (Py) underlayer can increase the magnetostatic coupling in lattices of nanomagnets with perpendicular magnetic anisotropy that do not show spontaneous ordering at experimental timescales[15]. The magnetostatic interactions in these systems are weak enough to prevent the individual nanomagnets from forming multidomain states but are also too weak to give spontaneous ordering. Thus, both the lack of rapid ordering among the nanomagnets and their tendency to remain in single-domain states are driven by the same principle of demagnetization field energy minimization. For example, increasing $D_{NM}$ provides stronger coupling but also, above a certain diameter, results in the undesirable formation of multidomain states. Thus, it is critical to increase the coupling energy in such a way that the demagnetization energy does not result in a multidomain state. To address this, we performed COMSOL calculations to determine how a Py underlayer would change the magnetic flux in a nanomagnet as well as through its neighbour. The purpose was to find parameters at which the magnetic flux through the nanomagnet itself is increased less significantly than the flux through its neighbour, thus promoting magnetostatic coupling more than formation of a multidomain state.

For the calculations, a saturation magnetization of 575 kA/m and 1063 kA/m was used for Py and Co, respectively. The magnetization was fixed to be in-plane for Py and out-of-plane for Co nanomagnets. The pair of circular Co nanomagnets were separated by 1 nm. The surroundings of the magnets were given the magnetic permeability of vacuum.

The dependence of the magnetic flux in the nanomagnet and a neighbouring nanomagnet (for a pair of coupled nanomagnets – see schematic in Fig. 10a) on the vertical separation of the Py layer from the bottom of the nanomagnets ("Py layer offset") is shown in Figures S10b and c. The ratio between the trends (Fig. S10d) indicates the Py layer offset where the flux through the neighbour is larger than that through the original nanomagnet. The best ratio of ~1.6 is achieved at a Py layer offset of 25 nm.

Using this as a guide, we fabricated lattices with $N_{Co}$ = 7, $t_{Co}$ = 1.40 nm and nanomagnet diameters $D_{NM}$ of 110, 140 and 170 nm on top of Py ($Ni_{80}Fe_{20}$) layers with thicknesses ranging from 0 to 7 nm and separated from the lattice by 25 nm. We only show the results for a Py thickness of 1 nm since this gave a noticeably better coupling enhancement than for the other thicknesses. The lattices were measured with MFM following the same procedure as described in the main text. At $D_{NM}$ = 170 nm, the majority of the nanomagnets were in a multidomain state, as expected due to the increased flux through the nanomagnet itself. The MFM measurement for $D_{NM}$ = 140 nm is shown in Figure S10f. Here, we observe a significant increase in spontaneous ordering compared to the system with no permalloy underlayer despite the larger $t_{Co}$ (1.45 nm compared with 1.40 nm) of the latter (Fig. S10e). Some multidomain states can also be seen in the nanomagnets in Figure S10f. Notably, significant spontaneous relaxation is also observed in the system with $D_{NM}$ = 110 nm, even for $S_{ASI}$ = 90 nm (Fig. S10g). At the same time, no multidomain states can be seen for these parameters.

These results show that enhancement of the intermagnet coupling by adding an underlayer can be significant enough to modify a system that is frozen on experimental timescales to one that relaxes to a low energy state. Nevertheless, special care must be taken to prevent the formation of multidomain states, which means that lattices with smaller $D_{NM}$ might benefit more from this approach.

**S11. Modification of $E_{MC}$ by altering the nanomagnet shape**

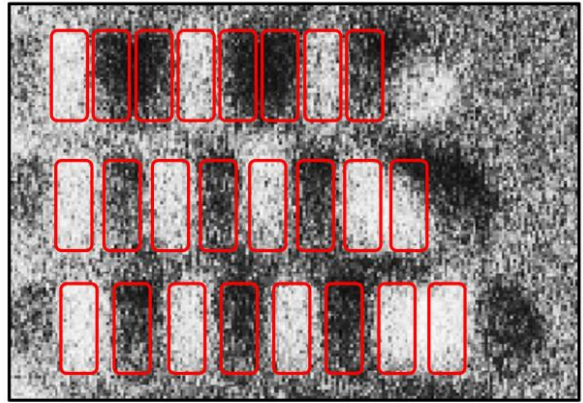

**Figure S11**. Chains of 112.5 nm × 49.5 nm nanomagnets with perpendicular magnetic anisotropy. The separations between the nanomagnets in the three rows, going from top to bottom, are 60, 50 and 40 nm.

Changing the geometry of a lattice and its nanomagnets can bring the "magnetic centres of mass" of the nanomagnets closer together and thus increase $E_{MC}$. To demonstrate this, we performed an MFM measurement of 1D chains of rectangular nanomagnets with rounded corners (Figure S11). While the ordering dynamics in such a system are not the same as in a 2D rectangular lattice, the degree of spontaneous ordering achieved at $t \sim 1000$ s for a given nanomagnet size can serve as a useful reference for the strength of the interaction between the nanomagnets. Here the same measurement protocol as in the main text was used. Almost perfect order is observed for a stack with $N_{Co}$ = 7, $t_{Co}$ = 1.4 nm despite separations of up to 60 nm and a much smaller nanomagnet area of ~5400 nm$^2$ compared with an area of ~22700 nm$^2$ for circular nanomagnets with $D_{NM}$ = 170 nm.

Such an approach could not only provide a way to enhance $E_{MC}$, as shown here, but also a method to tune the coupling strengths between different neighbours independently, by changing relative lengths of the borders between them. This fabrication of an array of nanomagnets with a precisely engineered spatially-varying intermagnet coupling would enable the creation of artificial spin ice with novel emergent properties.